\documentclass[12pt,preprint]{aastex}

\newcommand{\zabs}{$z_{\rm abs}$}
\newcommand{\zem}{$z_{\rm em}$}

\newcommand{\lya}{Ly$\alpha$}
\newcommand{\lyb}{Ly$\beta$}

\newcommand{\lyaf} {\lya\ forest}

\begin{document}

\title{The Kast Ground Based UV Spectral Survey of 79 QSOs 
at Redshift 2 for Lyman Alpha Forest and Metal Absorption}
\author{
David Tytler\altaffilmark{1,2},
John M. O'Meara\altaffilmark{1},
Nao Suzuki\altaffilmark{1},
David Kirkman\altaffilmark{1},
Dan Lubin\altaffilmark{1},
and Adam Orin\altaffilmark{1}
\altaffiltext{1}{Center for Astrophysics and Space Sciences,
University of California, San Diego,
MS 0424; La Jolla; CA 92093-0424}
\altaffiltext{2}{E-Mail: tytler@ucsd.edu}
}

\begin{abstract}
We present a moderate resolution ($\sim1.15$ \AA/pixel) 
survey of 79 quasars obtained using the Kast
spectrograph on the Shane 3m telescope at Lick observatory.  The
spectra span the wavelength range of 3175--5880 \AA, and have typical
signal to noise of 6--20 in the regions of the spectra showing \lyaf\ 
absorption.  The quasars have a mean emission redshift of
\zem$ =2.17$, and nearly all cover the entire \lyaf\ 
between \lya\ and \lyb.  Although
the quasars were selected to avoid BAL, two quasars in the survey
are BAL, one of which is a new discovery.
We list the H~I and metal ions observed in a total of 140
absorption systems.  We also identify 526
emission lines, and list their observed
wavelengths, along with new redshifts of the quasars.
We determine the rest wavelengths
of 3 emission lines or line blends in the forest to be 
$1070.95 \pm 1.00$, $1123.13 \pm 0.51$, and $1175.88 \pm 0.30$ \AA.
\end{abstract}


\section{Introduction}
In recent years, the ability to obtain precision cosmological
measurements from the high redshift intergalactic medium (IGM) has
been realized via the combination of large data-sets and numerical
simulations. 
Large QSO surveys such as the Sloan Digital Sky Survey
(SDSS\footnote{http://www.sdss.org}) and the Two Degree Field Survey 
(2dF\footnote{http://www.aao.gov.au/2df/})
have made available spectra of thousands of quasars sampling the universe in
\lyaf\ absorption at redshifts $z\simeq 2.5$ and larger.  

At redshifts
below 2.5, however, few moderately sized ground based surveys exist
which cover the \lyaf\ with sufficient resolution or
signal to noise.  Kim et al (2004) discuss the \lyaf\ from a sample of 27
high resolution VLT/UVES spectra with a median redshift of $
\left<z\right>=2.25$. Scott, Bechtold, and Dobrzycki (2000) present
$\sim 1$\AA\ resolution data for 39 QSOs targeted specifically to
cover the \lyaf\ at $1.6 < z < 2$.  The survey of Lyman limit
absorption by Sargent, Steidel, and Boksenberg (1989) contains 59
QSOs, some of which covering the \lyaf\ at $z<2.5$, but the resolution
is low, and the bulk of the data is at higher redshift.
Barthel, Tytler, and Thompson
(1990) present 67 spectra of radio loud quasars, of which 26 have
emission redshifts between 1.9 and 2.6, but this survey
contains no data below 3880 \AA. Sargent, Boksenberg, and Steidel
(1988) discuss $\sim 1.5$\AA\ resolution spectra of  55 QSOs, of 
which 33 have emission redshifts between
1.9 and 2.5, but this survey was targeted entirely for C~IV absorption,
and covered only wavelengths suitable for this metal transition.
Lanzetta, Wolfe, and Turnshek (1987) analyze 32 quasars for Mg~II absorption, 
16 of which have emission redshifts
between 1.9 and 2.5, but their survey was constrained to wavelengths
greater than 6200 \AA, and thus does not cover the \lyaf.
Finally, larger surveys such as the Large Bright Quasar Survey
(Hewett, Foltz, and Chaffee, 1995), the FIRST Bright QSO Survey (Gregg
et al, 1996) and the Hamburg/ESO survey for 
bright QSOs (Wisotzki et al., 2000) have a large number of QSOs below
$z=2.5$, but the resolution of these surveys is very low.

The scarcity of ground based data covering the \lyaf\ at redshifts 1.6--2.5
occurs primarily because working at observed wavelengths less than
4000 \AA\ becomes prohibitive due to issues such as poor fiber optic 
transmission, diminishing CCD response, and increased effects of 
atmospheric extinction. 

In this paper, we present a survey of 79 quasars with an average
emission redshift of \zem$=2.17$.  The redshift range of the \lyaf\ in
the survey serves as a compliment to larger surveys at higher
redshift, and is of sufficient size to provide a statistically
useful sampling of the IGM at a redshift of 2.  Various details of
the survey are also discussed in Tytler et al. (2004), where we
present a measurement and interpretation of the mean amount of
absorption due to H~I in the $z \sim 2$ intergalactic medium.
 
In section 2, we detail the observations and data reduction.  In
section 3, we present the spectra, and discuss individual quasars,
 metal absorption systems, and emission lines.

\section{Observations and Data Reduction}
The data in the survey were obtained using the Kast double spectrograph
on the Shane 3m telescope at Lick observatory over the years 2001 through 
2003.  In most cases, the spectra were obtained using the 2 arcsecond slit
width, although some observations required the 3 arcsecond  slit 
to accommodate poor seeing conditions, and in very rare cases, a 1.5 arcsecond
slit was used when the seeing conditions were very good.  
In all observations, the spectrograph slit was oriented 
such that it was aligned with the vertical direction at the effective middle 
of the exposure to minimize losses due to atmospheric dispersion. For
each QSO in the sample, exposures were taken using both the blue and
red cameras on Kast.  The blue camera used the 830/4360 grism, and the
red camera used the 1200/5000 grating, with a dispersion of 1.13 and
1.17 \AA\ per pixel respectively ($\sim 3$ pixels per resolution element).  
We employed the d46 dichroic to
split the light between the cameras.  The blue camera exposures covered
the approximate wavelength range of 3175--4540 \AA, and the red
camera exposures covered the approximate wavelength range 4475--5880
\AA.  Because of differences in the exact location of the blue camera
CCD on a given observing run, the starting and ending wavelengths 
varied by approximately $\pm$ 5 \AA.

The quasars in the survey were all chosen from the NED extragalactic database
\footnote{http://nedwww.ipac.caltech.edu}
with only the constraints that they have an emission redshift between 
\zem$ =1.9$ and \zem$ =2.4$, 
be of sufficient brightness to keep exposure times less than a few hours, 
and not show BAL absorption. The \zem\ constraint was made
so as to maximize coverage of the \lyaf\ at $z \simeq 1.9$, and the choice to 
neglect BAL QSOs was made to minimize contaminating absorption in
the spectra that do not come from the IGM.  We obtained and present spectra
for two QSOs, Q1542+5408 and Q2310+0018, which we found 
to show BAL absorption. Q1542+5408 is discussed in Green et al. (2001),
and Q2310+0018 is discovered to be BAL in this survey.
Other than these constraints,
the survey is unbiased with respect to the quasars observed.  

\begin{figure}
\epsscale{0.8}
\plotone{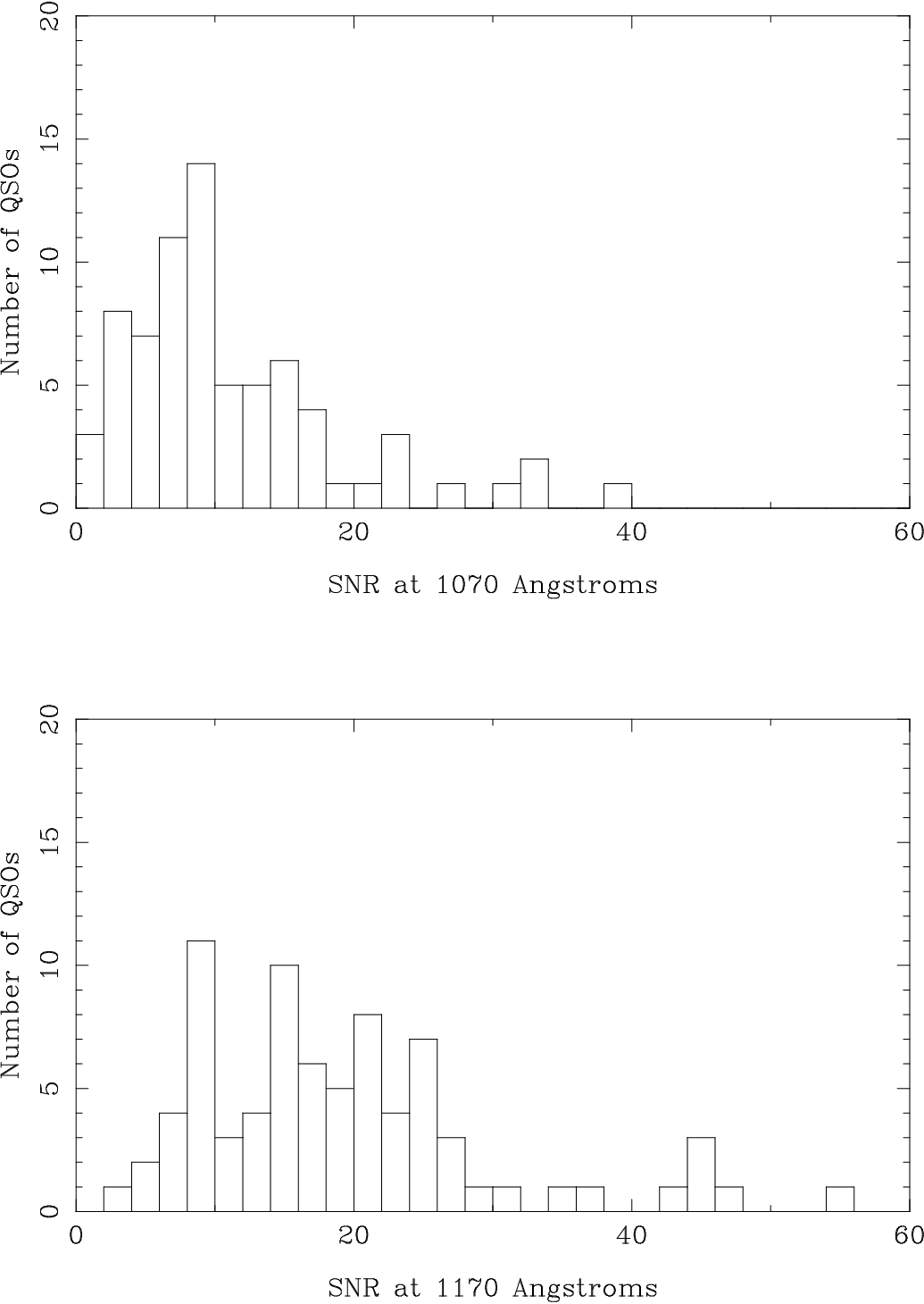}
\caption{Signal to noise distribution for the 79 quasars in the survey.  The
upper panel shows the signal to noise distribution for the data sampled over
20 \AA\ centered at 1070 \AA, while the lower panel shows the distribution
sampled over 20 \AA\ centered at 1170 \AA.}
\label{snrfig}
\end{figure}

Table \ref{obstab} lists the QSOs observed, along with their
B1950 and J2000 coordinates, the V magnitude as given by NED, the approximate 
redshift, and the
exposure time.  For simplicity, and to aid comparison with published spectra,
we choose to name the QSOs by
their abbreviated B1950 coordinates.  Table \ref{obstab2} lists the observation
date, slit width and signal to noise for each QSO.  The signal to noise
in table is calculated at two rest wavelengths, 1070\AA\ and 
1170\AA\ and is given as the mean signal to noise over 20 \AA\ centered about 
these wavelengths.  It should be noted that the true signal to noise of the 
spectrum will be in general higher, because absorption from the \lyaf\ lowers
the mean signal to noise.
These wavelengths were chosen because they correspond 
to the starting and ending
rest wavelengths that were used to sample the \lyaf\ for the measurement
of the mean flux decrement in Tytler et al. (2004).  Figure \ref{snrfig} shows
the distribution in signal to noise corresponding to the values in
Table \ref{obstab2}.

The exposure times were chosen with the 
intent to obtain S/N $> 10.0$ at wavelengths greater than the \lyb\
emission line of each QSO In practice, most of the spectra
reach this goal, except those observed in poor
conditions.  We do not present QSOs for which we obtained
S/N $<2$.  

We obtained, but do not present spectra for 6 objects
which upon reduction, are not QSOs.  
Two of these spectra are likely due to telescope
pointing error.  It is possible that the 4 other objects, Q1456+5404 
(14h56m47.71 +54d04m25.6 \zem = 2.300  V=16.50), Q1742+3749 
(17h42m 5.55 +37d49m08.3 \zem = 1.958  V=16.40), Q1755+5749 
(17h55m15.97 +57d49m06.9 \zem = 2.110  V=18.00), and Q2113+3004 
(21h13m59.42 +30d04m02.4 \zem = 2.080  V=17.30) are not QSOs.

All QSOs were reduced using the standard longslit reduction tools in
IRAF.  Wavelength calibration was also performed in the
standard manner with IRAF.  For each QSO, an arclamp observation was
made at the same telescope position as that for the QSO to minimize
the effects that flexure may have on the wavelength solution, and
typical wavelength errors were less than 1 \AA.  For
each night, a number of spectrophotometric flux standard stars were
observed for the purposes of flux calibration.  We list these stars in
Table \ref{fluxtab}.  As discussed in Suzuki et al. (2003), 
our errors in relative flux calibration could be as good as
a few percent.  Finally, the spectra were cleaned of any deviant
pixels resulting from poor sky or cosmic ray subtraction by replacing
these pixels with their neighboring flux values.

For the blue camera exposures, a continuum level was placed on the
spectrum by eye using the b-spline continuum fitting procedure discussed in
Kirkman  et.al (2003) and Tytler et. al (2004).  On average, in the regions
containing \lyaf\ absorption, the errors in this continuum level
placement were $< 5\%$.  A thorough discussion of the continuum level
error is given in Tytler et. al (2004).  The red camera exposures had
no continuum level assigned.

\section{The Kast $z\simeq 2$ Survey}
We now present the spectra of the survey, comment on absorbers in
individual QSOs, and tabulate the emission lines.

\subsection{Spectra}
\begin{figure}
\epsscale{0.8}
\plotone{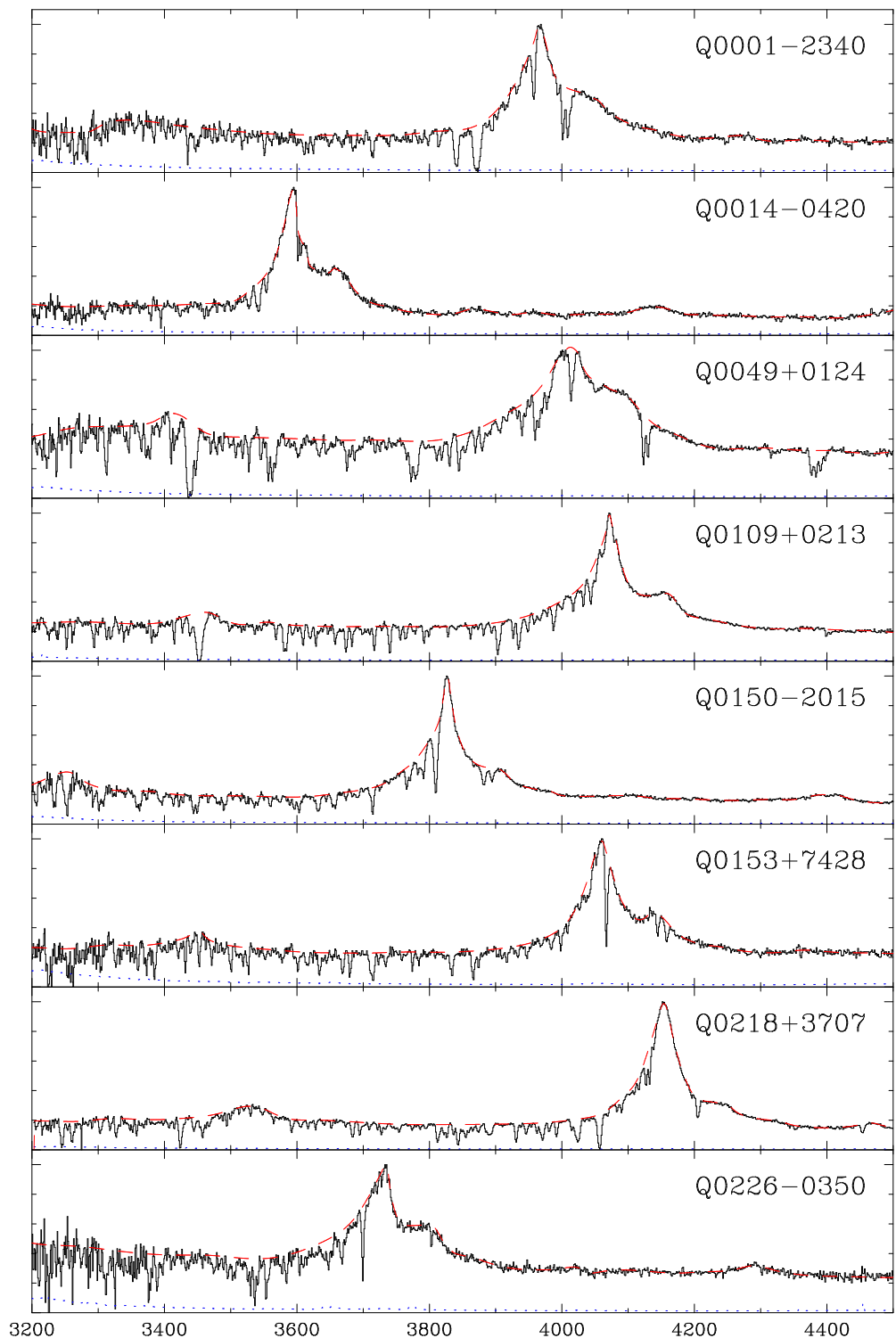}
\caption{Blue camera exposures of the QSOs in the Kast $z\simeq 2$
survey. For each QSO, the flux is shown
as the thick line, the 1$\sigma$ error in the flux as the dotted line, and the
dashed line represents the fit to the continuum level.  
The \textbf{y} axis is the flux level, 
in erg sec$^{-1}$ cm$^{-2}$ \AA$^{-1}$, on a linear scale with 
zero flux at the bottom, 
and the \textbf{x} axis is in units of \AA.}
\label{specfig}
\end{figure}

\begin{figure}
\epsscale{0.8}
\plotone{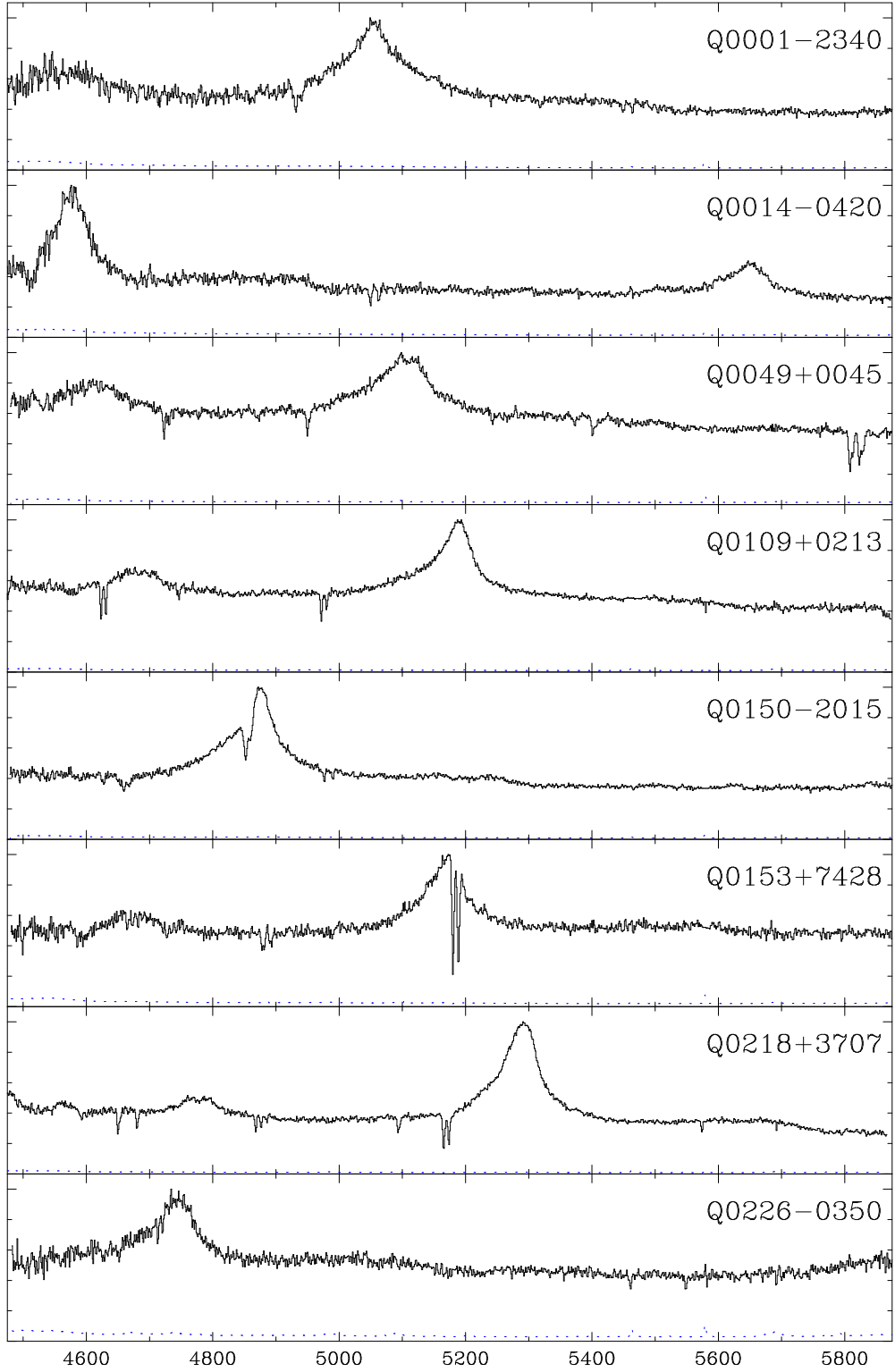}
\caption{Same as for Figure \ref{specfig}, but for the red camera.}

\label{redspecfig}
\end{figure}

\begin{figure}
\epsscale{0.8}
\plotone{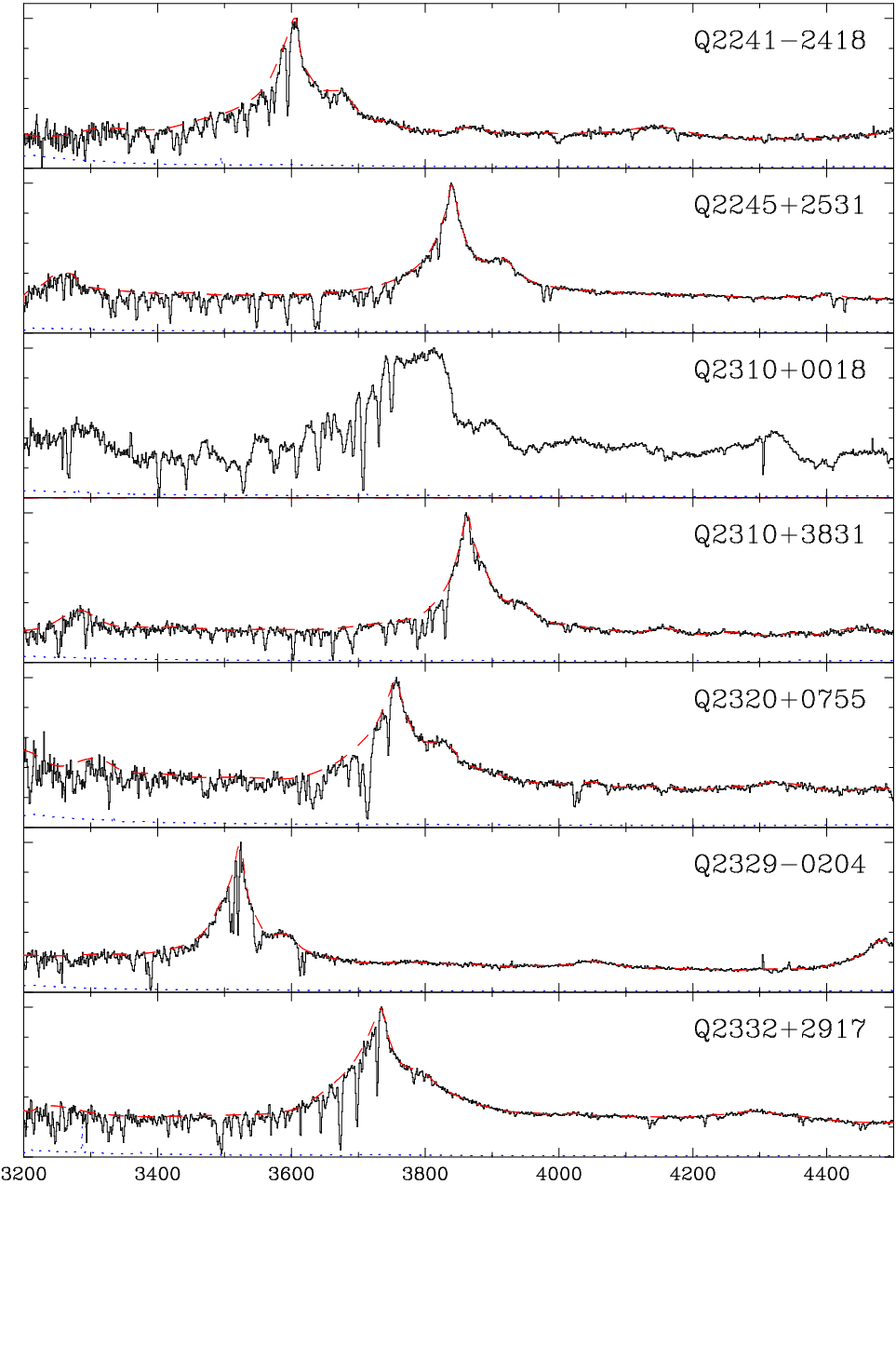}
\caption{Same as for Figure \ref{specfig}.}
\label{specfig10}
\end{figure}

\begin{figure}
\epsscale{0.8}
\plotone{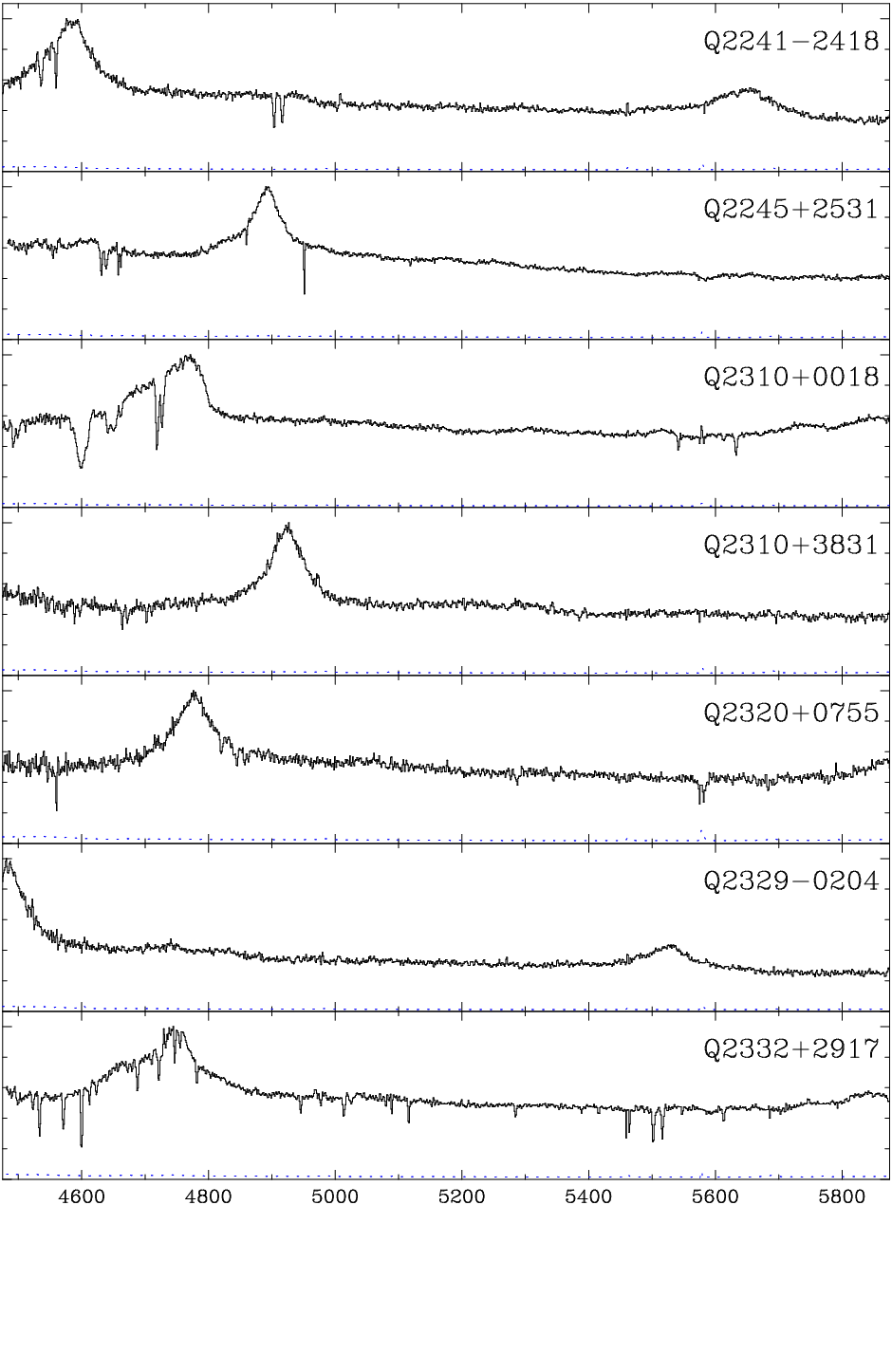}
\caption{Same as for Figure \ref{redspecfig}.}
\label{redspecfig10}
\end{figure}

In Figures \ref{specfig}--\ref{redspecfig10}, we show the blue and red camera
exposures of the 79 quasars which comprise the survey.  For each blue
exposure, the flux (solid line), error (dotted line), and continuum 
(dashed line) level is shown.  
For the red camera exposures, we display the flux and error levels.  
The spectra have well calibrated relative flux, 
but we do not assign a numerical
value to the flux, since absolute flux calibration was not a goal for
the survey, and few exposures were made under photometric conditions.

\subsection{Notes on Individual QSO Spectra}
We now discuss various aspects of the individual QSO spectra in the survey.  
Table
\ref{abstab} lists 140 systems with strong absorption in H~I and metals.
These systems were found by scanning the blue camera spectra by eye to find
systems which showed \lya\ absorption which appeared to have either 
significant
equivalent width, saturated absorption, or a combination of the two.  
For those systems, both the blue and red camera exposures 
were  then scanned for metal absorption at the same redshift as the H~I
absorber.  Finally, both blue and red camera exposures were inspected for
strong metal absorption.  When absorption was found, the ion
was identified, and the spectra were scanned for absorption from
other metal species at that redshift.  Spectral regions containing \lyaf\ 
absorption were not scanned for metal lines, as they would be blended
with H~I, and possibly misidentified.
In some cases, only a metal transition such as C~IV,
Mg~II, or Fe~II is given, since
the H~I at the redshift of these absorbers is either outside of the 
spectral coverage of the data, or is not visually strong compared to
the local \lyaf\ at nearby wavelengths.  

Table \ref{abstab} is not a complete census of
the metal absorption for the survey, but instead attempts to highlight systems
with a large H~I column density, such as Lyman limit and damped Lyman alpha 
systems, along with systems showing strong
absorption in metals.  Additional notes on individual QSOs are given 
below.

\textbf{Q1422+4224}: 
There appears to be strong H~I absorption at $z=1.951$.  Possible absorption
in metals at the expected positions of C~II and Si~IV is seen, but the SNR
at these wavelengths is poor.

\textbf{Q1542+3104}:
This QSOs contains considerable metal absorption near the 
C~IV emission line.  Moreover, the \lyaf\ in this QSOs appears to have
many regions of highly clustered absorption.  This QSO could be 
interpreted as a BAL (Weymann et al., 1991; Narayanan et al., 2004).

\textbf{Q1559+0853}:
The absorption from C~II corresponding at $z=2.252$ is uncertain, since the
Si~II absorption from the system at $z=1.842$ places Si~II at nearly the
same wavelengths.

\textbf{Q1649+4007}:
The Fe~II transitions for the absorber at $z=0.499$ which shows Mg~II are all
in the \lyaf, and are thus uncertain.

\subsection{Emission Lines}
Large samples of quasar emission line wavelengths have been presented and 
discussed by many authors (Wills, Netzer and Wills, 1985; Weymann et al., 1991;
Tytler and Fan, 1992; Laor et al. 1995; Forster et al. 2001; 
Constantin et al., 2002 and many others).  For this survey, we also present
the observed emission line wavelengths.

For each quasar in the survey, we attempted to identify all emission lines
with wavelengths greater than the O~VI--\lyb\ emission line blend.  The results of the
emission line identifications are given in Table \ref{emltab}, where the
observed wavelengths for each emission line are given.  In total, 526
emission lines were identified.
The wavelengths in Table \ref{emltab} were determined by either fitting
gaussians to the top third of the flux of the
lines when possible, or by determining the line
emission peak by eye.  For those emission lines in the \lyaf, all
wavelengths listed were determined by eye.  Because the methods used
to determine the emission line peaks are fairly inexact, and
because the effects of signal to noise and \lyaf\ and strong metal 
absorption hamper the process, we estimate that the peak wavelengths
have errors of $\pm 5$ \AA.

In the process of fitting
the quasar continuum level, as noted in Tytler et al. (2004), it was found
that emission line features were often required at rest wavelengths
of 1071, 1123, and 1176 \AA.  
The values for the rest wavelength,
$\lambda_{rest}$, listed in Table \ref{emltab} come
either from Table 2 of Telfer et al. (2002) or from Wills et
al. (1995), except for the lines near 1071, 1123, and 1176 \AA\ that we 
discuss below.  

In Table \ref{emlztab}, we list the redshifts, denoted by $z_{eff}$,
 of each quasar as
determined by taking the mean of redshifts given by each line
observed.  We do not use the lines in the first
three transitions listed in Table \ref{emltab}, because the errors on
these lines are in general larger.  To help determine
the exact rest wavelengths of the emission lines with wavelengths less than 
\lya, Table \ref{emlztab} also provides the rest wavelengths for the first
three transitions of Table \ref{emltab} as determined by dividing their
observed wavelengths by $(1 + z_{eff})$.  We find the rest wavelengths 
for these transitions to be $1070.95 \pm 1.00$, $1123.13 \pm 0.51$, and
$1175.88 \pm 0.30$ \AA.  These wavelengths are consistent with the 
identifications cited in Tytler et al. (2004).

In some cases, we note that the blend of O~I and Si~II at
$\lambda_{rest} = 1306$ \AA\ is dominated by the Si~II 1309 transition,
and may cause additional errors in the determination of the quasar's
redshift.  In Table \ref{emlextras}, we list those emission lines which
were observed in some spectra but are not in Table \ref{emltab}, and where
possible, identify the transition responsible for the line.
Finally, many quasars in the survey show a shelf like emission 
feature extending between He~II 1640 and O~III$]$.
We end with some notes on the emission lines for individual
quasars.

\textbf{Q0743+6601 and Q1122-1648:} The C~III$^{*}$ emission 
line is seen in this
QSO but is too heavily blended with the wing of \lya\ to
give a value for the peak position.

\textbf{Q2134+1531:} A broad emission shelf feature
is seen, consistent with the feature at $\lambda_{rest} \sim 1065$\AA\ blending
with the O~VI--\lyb\ blend.

\textbf{Q2140+2403:} This is the only QSO for which
we do not provide a measurement of the \lya\ emission line, due to the effects
of the very strong associated absorber near the expected emission line peak.

\textbf{Q2245+2531:} The Si~IV--O~IV$]$ blend
appears to be dominated entirely by Si~IV 1393.

\section{Acknowledgements}
This work was funded in part by grant NAG5-13113 from NASA and by
grant AST-0098731 from the NSF.  The spectra were obtained from the Lick
observatory and we thank the Lick observatory staff.  This research has
made extensive use of the NASA/IPAC Extragalactic Database (NED) which is
operated by JPL, under contract with NASA.


\clearpage

\begin{deluxetable}{lllcccl}
\tablewidth{0pt}
\tablecaption{\label{obstab}Kast $z\simeq 2$ Spectra of 79 QSOs}
\tablehead{
\colhead{Name} &
\colhead{Coordinates} &
\colhead{Coordinates} &
\colhead{V} &
\colhead{\zem\tablenotemark{A}} &
\colhead{Exposure}
\\
\colhead{} &
\colhead{} &
\colhead{} &
\colhead{} &
\colhead{} &
\colhead{Time} 
\\
\colhead{(1950)} &
\colhead{(B1950)} &
\colhead{(J2000)} &
\colhead{} &
\colhead{} &
\colhead{(seconds)} 
}
\startdata
Q0001--2340 & 00 01 11.50 --23 40 37.0  & 00 03 44.95 --23 23 54.7 & 16.70  & 2.26 &  1380 \\
Q0014--0420 & 00 14 09.36 --04 20 57.0  & 00 16 42.80 --04 04 17.0 & 16.65 & 1.96 &  720   \\
Q0049+0124 & 00 49 59.56 +01 24 23.3 & 00 52 33.71 +01 40 40.5 & 17.00 & 2.29 &  2700   \\
Q0109+0213 & 01 09 42.31 +02 13 53.1 & 01 12 16.91 +02 29 47.6 & 17.64 & 2.34 &  4800   \\
Q0150--2015 & 01 50 04.98 --20 15 52.6 & 01 52 27.30 --20 01 06.0 & 17.10 & 2.14 &  2700   \\
Q0153+7428 & 01 53 04.33 +74 28 05.6 & 01 57 34.96 +74 42 43.2 &  16.00 & 2.34 &  1800   \\
Q0218+3707 & 02 18 02.70 +37 07 04.1 & 02 21 05.50 +37 20 46.0 & 17.50 & 2.41 &  3600   \\
Q0226--0350 & 02 26 22.08 --03 50 58.6 & 02 28 53.21 --03 37 37.1 & 16.96 & 2.07 &  2400   \\
Q0248+3402 & 02 48 23.35 +34 02 22.0 & 02 51 27.70 +34 14 41.0 & 17.70 & 2.22 &  4200   \\
Q0348+0610 & 03 48 36.62 +06 10 15.5 & 03 51 16.53 +06 19 14.2 & 17.60 & 2.05 &  3600   \\
Q0421+0157 & 04 21 32.67 +01 57 32.7 & 04 24 08.56 +02 04 24.9 & 17.04 & 2.04 &  3000   \\
Q0424--1309 & 04 24 47.81 --13 09 32.9 & 04 27 07.30 --13 02 53.0 & 17.50 & 2.16 &  4200   \\
Q0450--1310 & 04 50 54.00 --13 10 39.0 & 04 53 12.83 --13 05 46.1 & 16.50 & 2.25 &  3600   \\
Q0726+2531 & 07 26 25.23 +25 31 07.2 & 07 29 28.47 +25 24 51.9 & 17.81 & 2.30 &  4800   \\
Q0743+6601 & 07 43 58.60 +66 01 00.0 & 07 48 46.22 +65 53 31.4 & 17.00 & 2.20 &  3600   \\
Q0748+6105 & 07 48 01.84 +61 05 35.9 & 07 52 22.50  +60 57 52.0 & 17.50 & 2.49 &  4200   \\
Q0752+3429 & 07 52 10.00 +34 29 31.3 & 07 55 24.00 +34 21 34.0 & 17.80 & 2.12 &  4500   \\
Q0800+3031 & 08 00 34.43 +30 31 24.1 & 08 03 42.00 +30 22 55.0 & 16.70 & 2.02 &  3600   \\
Q0836+7104 & 08 36 21.53 +71 04 22.5 & 08 41 24.36 +70 53 42.2 & 16.50 & 2.18 &  2400   \\
Q0854+3324 & 08 54 21.61 +33 24 53.0 & 08 57 26.95 +33 13 17.2 & 17.43 & 2.33 &  4200   \\
Q0907+3811 & 09 07 44.95 +38 11 31.9 & 09 10 54.20  +37 59 15.0 & 17.30 & 2.15 &  3000   \\
Q0936+3653 & 09 36 32.36 +36 53 35.9 & 09 39 35.10  +36 40 00.0 & 17.00 & 2.02 &  2400   \\
Q0937--1818 & 09 37 30.22 --18 18 37.3 & 09 39 51.10 --18 32 15.0 & 16.20 & 2.36 &  3000   \\
Q1023+3009 & 10 23 58.86 +30 09 29.9 & 10 26 48.10 +29 54 12.0 & 17.10 & 2.33 &  3600   \\
Q1103+6416 & 11 03 03.98 +64 16 21.9 & 11 06 10.70 +64 00 09.0 & 15.80 & 2.20 &  1800   \\
Q1116+2106 & 11 16 44.49 +21 06 11.4 & 11 19 22.90 +20 49 46.0 & 17.30 & 2.46 &  3600   \\
Q1122--1648 & 11 22 12.28 --16 48 47.4 & 11 24 42.80 --17 05 17.0 & 16.50 & 2.40 &  2700  \\
Q1130+3135 & 11 30 06.36 +31 35 24.3 & 11 32 45.20 +31 18 50.0 & 17.00 & 2.29 &  2400   \\
Q1147+6556 & 11 47 53.77 +65 56 08.8 & 11 50 34.50 +65 39 28.0 & 16.20 & 2.21 &  2360   \\
Q1222+2251 & 12 22 56.58 +22 51 49.3 & 12 25 27.40 +22 35 13.0 & 15.49 & 2.05 &  3000   \\
Q1224--0812 & 12 24 02.65 --08 12 52.8 & 12 26 37.50 --08 29 29.0 & 16.83 & 2.16 &  2400   \\
Q1224+2905 & 12 24 57.90 +29 05 23.0 & 12 27 27.40 +28 48 47.0 & 17.00 & 2.25 &  3600   \\
Q1225+3145 & 12 25 56.07 +31 45 12.6 & 12 28 24.96 +31 28 37.6 & 15.87 & 2.18 &  2100   \\
Q1231+2924 & 12 31 27.08 +29 24 20.8 & 12 33 55.51 +29 07 48.9 & 16.84 & 2.01 &  2400   \\
Q1247+2657 & 12 47 39.09 +26 47 27.1 & 12 50 05.75 +26 31 07.7 & 15.80 & 2.03 &  1200   \\
Q1251+2636 & 12 51 56.97 +26 36 21.8 & 12 54 23.08 +26 20 06.5 & 16.45 & 2.03 &  2100   \\
Q1307+4617 & 13 07 58.49 +46 17 20.8 & 13 10 11.60 +46 01 24.0 & 16.74 & 2.13 &  3000   \\
Q1312+7837 & 13 12 30.24 +78 37 44.6 & 13 13 21.30 +78 21 53.0 & 16.40 & 2.00 &  2400   \\
Q1326+3923 & 13 26 10.24 +39 23 47.2 & 13 28 23.70 +39 08 17.0 & 16.60 & 2.32 &  2100   \\
Q1329+4117 & 13 29 29.82 +41 17 22.7 & 13 31 41.10 +41 01 58.0 & 16.30 & 1.93 &  4500  \\
Q1331+1704 & 13 31 10.00 +17 04 25.8 & 13 33 35.78 +16 49 04.0 & 16.71 & 2.08 &  2700   \\
Q1416+0906 & 14 16 23.30 +09 06 14.0 & 14 18 51.09 +08 52 27.1 & 17.00 & 2.01 &  2400   \\
Q1418+2254 & 14 18 51.06 +22 54 58.3 & 14 21 08.72 +22 41 17.4 & 16.60 & 2.19 &  1800   \\
Q1422+4224 & 14 22 37.86 +42 24 01.6 & 14 24 36.00 +42 10 30.0 & 17.10 & 2.21 &  4200   \\
Q1425--1338 & 14 25 02.87 --13 38 19.4 & 14 27 46.40 --13 51 44.0 & 17.19 & 2.03 &  2400   \\
Q1435+6349 & 14 35 37.25 +63 49 36.0 & 14 36 45.80 +63 36 37.8 & 15.00 & 2.06 &  1500   \\
Q1517+2556 & 15 17 08.11 +23 56 52.0 & 15 19 19.40 +23 46 02.0 & 16.40 & 1.90 &  4500 \\
Q1542+3104 & 15 42 48.48 +31 04 42.0 & 15 44 49.00 +30 55 21.0 & 17.00 & 2.28 &  2400   \\
Q1542+5408\tablenotemark{B} & 15 42 41.88 +54 08 25.6 & 15 43 59.40 +53 59 03.0 & 16.00 & 2.36 &  1800   \\
Q1559+0853 & 15 59 57.80 +08 53 53.0 & 16 02 22.56 +08 45 36.3 & 16.70 & 2.26 &  2400   \\
Q1611+4719 & 16 11 10.53 +47 19 32.2 & 16 12 39.90 +47 11 57.0 & 17.60 & 2.38 &  3600   \\
Q1618+5303 & 16 18 28.71 +53 03 20.0 & 16 19 42.30 +52 56 13.0 & 17.50 & 2.34 &  4500   \\
Q1626+6433 & 16 26 20.41 +64 33 32.3 & 16 26 45.60 +64 26 55.0 & 15.80 & 2.31 &  1800   \\
Q1632+3209 & 16 32 17.96 +32 09 45.7 & 16 34 12.78 +32 03 35.4 & 16.93 & 2.34 &  2100   \\
Q1649+4007 & 16 49 57.45 +40 07 16.7 & 16 51 37.56 +40 02 18.7 & 17.18 & 2.33 &  2400   \\
Q1703+5350 & 17 03 01.49 +53 50 57.2 & 17 04 06.70 +53 46 53.0 & 17.40 & 2.37 &  3600   \\
Q1705+7101 & 17 05 00.60 +71 01 34.0 & 17 04 26.08 +70 57 34.7 & 17.50 & 2.01 &  4200   \\
Q1716+4619 & 17 16 01.69 +46 19 39.0 & 17 17 26.80 +46 16 31.0 & 17.10 & 2.11 &  2400   \\
Q1720+2501 & 17 20 49.93 +25 01 20.6 & 17 22 52.99 +24 58 34.7 & 17.10 & 2.25 &  4000  \\
Q1754+3818 & 17 54 58.67 +38 18 10.2 & 17 56 39.70 +38 17 52.0 & 17.50 & 2.16 &  2700   \\
Q1833+5811 & 18 33 09.83 +58 11 07.8 & 18 33 57.00 +58 13 34.0 & 17.00 & 2.03 &  1800   \\
Q1834+6117 & 18 34 46.51 +61 17 07.3 & 18 35 19.68 +61 19 40.0 & 17.60 & 2.27 &  2700   \\
Q1848+6705 & 18 48 26.30 +67 05 07.0 & 18 48 25.35 +67 08 37.2 & 17.50 & 2.03 &  1500   \\
Q2044--1650 & 20 44 30.78 --16 50 09.4 & 20 47 19.67 --16 39 05.8 & 17.36 & 1.94 &  3600   \\
Q2103+1843 & 21 03 50.50 +18 43 46.0 & 21 06 08.52 +18 55 49.9 & 16.80 & 2.21 &  2400   \\
Q2134+0028 & 21 34 05.21 +00 28 25.0 & 21 36 38.59 +00 41 54.2 & 16.79 & 1.94 &  2400   \\
Q2134+1531 & 21 34 01.07 +15 31 38.2 & 21 36 23.80 +15 45 07.0 & 17.30 & 2.13 &  3600   \\
Q2135+1326 & 21 35 40.84 +13 26 19.9 & 21 38 05.20 +13 39 53.0 & 17.10 & 2.30 &  2700   \\
Q2140+2403 & 21 40 31.75 +24 03 33.0 & 21 42 48.50 +24 17 18.0 & 16.80 & 2.16 &  2400   \\
Q2147--0825 & 21 47 09.10 --08 25 17.9 & 21 49 48.17 --08 11 16.2 & 16.18 & 2.12 &  2400  \\
Q2150+0522 & 21 50 54.30 +05 22 08.6  & 21 53 24.67 +05 36 18.9 & 17.77 & 1.98 &  6000   \\
Q2157--0036 & 21 57 20.39 --00 36 15.3 & 21 59 54.45 --00 21 50.3 & 16.98 & 1.96 &  4200   \\
Q2241--2418 & 22 41 56.68 --24 18 48.8 & 22 44 40.30 --24 03 02.0 & 16.95 & 1.96 &  1800   \\
Q2245+2531 & 22 45 03.78 +25 31 39.3 & 22 47 27.40 +25 47 30.0 & 17.60 & 2.16 &  6000   \\
Q2310+0018\tablenotemark{B} & 23 10 50.80 +00 18 24.1 & 23 13 24.45 +00 34 44.5 & 17.00 & 2.08 & 1800 \\
Q2310+3831 & 23 10 36.18 +38 31 22.7 & 23 12 58.79 +38 47 42.6 & 17.50 & 2.18 &  3000   \\
Q2320+0755 & 23 20 03.91 +07 55 33.6 & 23 22 36.08 +08 12 01.6 & 17.50 & 2.08 &  4500   \\
Q2329-0204 & 23 29 02.27 --02 04 40.4 & 23 31 36.33 --01 48 06.5 & 17.00 & 1.89 &  2400   \\
Q2332+2917 & 23 32 32.03 +29 17 39.5 & 23 35 01.50 +29 34 15.0 & 17.60 & 2.07 &  6000   \\
\enddata
\tablenotetext{A}{\footnotesize{This redshift is approximate, from the 
\lya\ peak}}
\tablenotetext{B}{\footnotesize{BAL QSO}}
\end{deluxetable}

\begin{deluxetable}{lccccl}
\tablewidth{0pt}
\tablecaption{\label{obstab2}Kast $z\simeq 2$ Spectra of 79 QSOs}
\tablehead{
\colhead{Name} &
\colhead{Observation Date} &
\colhead{Slit Width} &
\colhead{SNR} &
\colhead{SNR} 
\\
\colhead{(1950)} &
\colhead{(Year-Month-Day)} &
\colhead{(arcseconds)} &
\colhead{(1070 \AA)} &
\colhead{(1170 \AA)} 
}
\startdata
Q0001--2340 & 2001-07-20  &   3.0  & 9.04 & 14.33  \\
Q0014--0420 & 2001-07-19  &   3.0  & 3.19 & 8.88   \\
Q0049+0124  & 2001-09-13  &   3.0  & 15.24 & 24.00 \\
Q0109+0213  & 2001-12-16  &   2.0  & 22.73 & 34.25 \\
Q0150--2015 & 2001-09-13  &   3.0  & 9.26 & 19.29  \\
Q0153+7428  & 2001-09-13  &   3.0  & 7.53 & 13.14  \\
Q0218+3707  & 2002-01-11  &   2.0  & 22.08 & 29.59 \\
Q0226--0350 & 2003-01-30  &   2.0  & 5.45 & 14.12  \\
Q0248+3402  & 2002-01-11  &   2.0  & 23.27 & 36.34 \\
Q0348+0610  & 2003-01-30  &   2.0  & 6.02 & 16.54  \\
Q0421+0157  & 2001-12-16  &   2.0  & 6.19 & 16.09  \\
Q0424--1309 & 2002-01-11  &   2.0  & 1.71 & 2.95   \\
Q0450--1310 & 2003-01-29  &   2.0  & 11.99 & 19.20 \\
Q0726+2531  & 2003-01-31  &   2.0  & 8.94 & 14.36  \\
Q0743+6601  & 2001-12-16  &   2.0  & 17.36 & 25.15 \\
Q0748+6105  & 2001-12-16  &   2.0  & 16.62 & 20.73 \\
Q0752+3429  & 2003-01-30  &   2.0  & 9.18 & 15.98  \\
Q0800+3031  & 2002-04-09  &   1.5  & 8.09 & 21.69  \\
Q0836+7104  & 2002-03-09  &   2.0  & 15.92 & 23.20 \\
Q0854+3324  & 2003-01-30  &   2.0  & 7.29 & 11.77  \\
Q0907+3811  & 2003-01-31  &   2.0  & 5.42 & 9.70   \\
Q0936+3653  & 2003-01-31  &   2.0  & 1.24 & 5.50   \\
Q0937--1818 & 2003-01-29  &   2.0  & 9.59 & 15.87  \\
Q1023+3009  & 2002-04-09  &   1.5  & 6.28 & 9.02   \\
Q1103+6416  & 2003-01-31  &   2.0  & 15.16 & 24.08 \\
Q1116+2106  & 2002-04-11  &   1.5  & 4.44 & 5.64   \\
Q1122--1648 & 2003-01-29  &   2.0  & 15.23 & 25.84 \\
Q1130+3135  & 2003-01-30  &   2.0  & 9.69 & 15.89  \\
Q1147+6556  & 2002-04-09  &   1.5  & 3.54 & 6.55   \\
Q1222+2251  & 2003-01-29  &   2.0  & 9.24 & 22.63  \\
Q1224--0812 & 2003-01-31  &   2.0  & 6.48 & 12.82  \\
Q1224+2905  & 2002-04-11  &   1.5  & 4.85 & 6.54   \\
Q1225+3145  & 2001-05-19  &   3.0  & 27.77 & 45.87 \\
Q1231+2924  & 2003-01-30  &   2.0  & 10.16 & 22.86 \\
Q1247+2657  & 2001-05-19  &   2.0  & 21.49 & 45.37 \\
Q1251+2636  & 2001-05-19  &   2.0  & 5.28 & 15.57  \\
Q1307+4617  & 2002-03-09  &   2.0  & 13.89 & 24.74 \\
Q1312+7837  & 2003-01-29  &   2.0  & 8.36 & 20.90  \\
Q1326+3923  & 2003-01-31  &   2.0  & 11.63 & 16.66 \\
Q1329+4117  & 2003-07-28  &   2.0  & -- \tablenotemark{A} & 9.49 \\
Q1331+1704  & 2002-04-11  &   1.5  & 7.53 & 14.35  \\
Q1416+0906  & 2003-01-29  &   2.0  & 3.37 & 10.42  \\
Q1418+2254  & 2001-05-19  &   2.0  & 11.98 & 20.80 \\
Q1422+4224  & 2002-03-09  &   2.0  & 5.22 & 9.53   \\
Q1425--1338 & 2003-01-30  &   2.0  & 1.50 & 6.34   \\
Q1435+6349  & 2001-05-19  &   2.0  & 8.14 & 18.13  \\
Q1517+2556  & 2003-07-29  &   2.0  & -- \tablenotemark{A} & 8.31 \\
Q1542+3104  & 2001-05-18  &   2.0  & 7.83 & 9.60   \\
Q1542+5408\tablenotemark{B}  & 2001-05-19  &   2.0  & 17.20 & 21.52 \\
Q1559+0853  & 2001-07-19  &   3.0  & 12.65 & 21.96 \\
Q1611+4719  & 2002-04-11  &   1.5  & 8.94 & 16.60  \\
Q1618+5303  & 2001-07-20  &   3.0  & 39.17 & 54.79 \\
Q1626+6433  & 2001-07-18  &   3.0  & 30.55 & 45.82 \\
Q1632+3209  & 2001-05-19  &   2.0  & 9.99 & 16.28  \\
Q1649+4007  & 2001-07-18  &   3.0  & 18.98 & 27.32 \\
Q1703+5350  & 2001-07-18  &   3.0  & 32.88 & 43.43 \\
Q1705+7101  & 2001-07-19  &   2.0  & 2.74 & 8.52   \\
Q1716+4619  & 2001-05-18  &   3.0  & 13.29 & 19.53 \\
Q1720+2501  & 2001-07-17  &   3.0  & 12.49 & 18.50 \\
Q1754+3818  & 2001-05-19  &   2.0  & 7.90 & 26.11  \\
Q1833+5811  & 2001-05-18  &   3.0  & 3.41 & 8.42   \\
Q1834+6117  & 2001-05-19  &   2.0  & 9.29 & 14.67  \\
Q1848+6705  & 2001-05-18  &   3.0  & 3.71 & 8.38   \\
Q2044--1650 & 2001-07-18  &   3.0  & -- \tablenotemark{A} & 15.63 \\
Q2103+1843  & 2001-07-17  &   3.0  & 32.90 & 46.93 \\
Q2134+0028  & 2001-07-19  &   3.0  & -- \tablenotemark{A} & 22.77 \\
Q2134+1531  & 2001-07-18  &   3.0  & 15.16 & 25.12 \\
Q2135+1326  & 2001-07-17  &   3.0  & 15.78 & 24.27 \\
Q2140+2403  & 2001-07-17  &   3.0  & 17.08 & 31.21 \\
Q2147--0825 & 2003-07-28  &   2.0  & 4.36 & 9.49   \\
Q2150+0522  & 2003-07-29  &   2.0  & 3.97 & 7.96   \\
Q2157--0036 & 2003-07-28  &   2.0  & -- \tablenotemark{A} & 20.79 \\
Q2241--2418 & 2001-07-17  &   3.0  & 2.69 & 10.95  \\
Q2245+2531  & 2003-07-28  &   2.0  & 13.23 & 26.09 \\
Q2310+0018\tablenotemark{B}  & 2001-07-11  & 3.0 & 11.71 & 20.45 \\
Q2310+3831  & 2001-07-18  &   3.0  & 10.80 & 17.77 \\
Q2320+0755  & 2001-07-20  &   3.0  & 6.38 & 12.66  \\
Q2329-0204  & 2001-07-19  &   3.0  & 6.38 & 12.66  \\
Q2332+2917  & 2003-07-29  &   2.0  & 8.43 & 21.88  \\
\enddata
\tablenotetext{A}{\footnotesize{This rest wavelength was not covered by 
this spectrum}}
\tablenotetext{B}{\footnotesize{BAL QSO}}
\end{deluxetable}

\clearpage
\begin{deluxetable}{lccc}
\tablewidth{0pt}
\tablecaption{\label{fluxtab}Standard Stars for Flux Calibration}
\tablehead{
\colhead{Name} &
\colhead{Coordinates (J2000)} &
\colhead{V} &
\colhead{Spectral Type}
}
\startdata
G191B2B      & 05 05 30.6 + 52 49 54  & 11.78 & DAO \\
Feige 34     & 10 39 36.7 + 43 06 10  & 11.18 & DO \\
Feige 67     & 12 41 51.8 + 17 31 20  & 11.81 & sdO \\
BD+33d2642   & 15 51 59.9 + 32 56 55  & 10.81 & B2IV \\
BD+28d4211   & 21 51 11.1 + 28 51 52  & 10.51 & 0p \\
Feige 110    & 23 19 58.4 -- 05 09 56 & 11.82 & DOp \\
\enddata
\end{deluxetable}

\clearpage
\begin{deluxetable}{llll}
\tablewidth{0pt}
\tablecaption{\label{abstab}Strong Absorption Systems}
\tablehead{
\colhead{Name} &
\colhead{\zabs} &
\colhead{Ions} &
\colhead{Notes}
\\
\colhead{(1950)} &
\colhead{} &
\colhead{} &
\colhead{}
}
\startdata
Q0001--2340 & 2.184 & H~I, C~II, C~IV, Al~II, Si~II, Si~IV & \\
            & 0.431 & Mg~II                   & \\
Q0014--0420 & 0.806 & Mg~II						 & \\
Q0049+0124  & 1.828 & H~I, C~IV, Al~II, Al~III, Si~IV & \\
				& 1.663 & C~IV                    & \\
				& 1.077 & Mg~II, Fe~II				 & \\
Q0109+0213  & 2.212 & C~IV							 & \\
				& 1.987 & C~IV							 & \\
				& 1.841 & H~I, C~IV, Al~II			 & \\
Q0150--2015 & 2.134 & H~I, C~IV, N~V, Si~IV   & \tablenotemark{A} \\
				& 2.010 & C~IV							 & \\
				& 0.780 & Mg~II, Fe~II            & \\
Q0153+7428  & 2.346 & H~I, C~IV, N~V			 & \tablenotemark{A} \\
	 			& 0.745 & Mg~II						 & \\ 
Q0218+3707  & 2.337 & H~I, C~II, C~IV, O~I, Al~II, Si~II, Si~IV & \\
				& 2.144 & C~IV							 & \\
Q0248+3402  & 2.167 & H~I, C~IV               & \\
Q0348+0610  & 2.034 & H~I, C~IV, Si~IV			 & \\
 				& 2.026 & H~I, C~IV, Si~IV			 & \\
			   & 1.969 & H~I, C~IV, Si~IV			 & \\
Q0421+0157  & 1.637 & H~I, C~IV					 & \\
Q0450--1310 & 2.231 & H~I, C~IV, N~IV         & \tablenotemark{A} \\
				& 2.107 & C~IV, Si~IV             & \\
            & 2.067 & H~I, C~II, C~IV, O~I, Si~II, Si~IV & \tablenotemark{B} \\
				& 0.494 & Mg~II                   & \\
Q0726+2531  & 2.288 & H~I, C~IV, N~V, O~VI    & \tablenotemark{A} \\
Q0743+6601  & 1.695 & H~I, C~IV					 & \\
				& 1.650 & H~I, C~IV					 & \\
				& 0.638 & Mg~II 						 & \\
Q0748+6105  & 2.209 & H~I, C~IV					 & \\
Q0752+3429  & 2.068 & C~IV							 & \\
				& 2.051 & H~I							 & \tablenotemark{C}\\
				& 1.501 & C~IV							 & \\
			   & 1.063 & Mg~II, Fe~II 				 & \\
Q0800+3031  & 1.862 & H~I							 & \tablenotemark{C}\\
Q0854+3324  & 2.269 & H~I, C~II, C~IV, O~I, Al~II, Si~II, Si~IV   & \\
Q0836+7104  & 0.915 & Mg~II, Fe~II				 & \\
Q0907+3811  & 1.820 & H~I, C~IV               & \\
Q0937--1818 & 0.941 & Mg~II, Fe~II				 & \\
Q1103+6416  & 1.941 & C~IV							 & \\
				& 1.892 & H~I, Si~IV              & \\
Q1122--1648 & 0.682 & Mg~II, Fe~II				 & \\
Q1130+3135  & 2.210 & C~IV							 & \\
				& 2.112 & C~IV							 & \\
				& 2.021 & H~I, C~II, C~IV, Al~II, Si~II, Si~IV & \\
Q1222+2251  & 1.936 & H~I, C~II, C~IV, Si~IV  & \\
				& 1.486 & C~IV							 & \\
				& 0.668 & Mg~II						 & \\
Q1224--0812 & 1.890 & H~I, C~IV					 & \\
				& 1.602 & C~IV, Al~II, Al~III, Si~II & \\
Q1225+3145  & 2.121 & C~IV							 & \\
				& 1.795 & H~I, C~IV, Al~II, Al~III, Si~II, Si~IV & \\
				& 1.626 & C~IV							 & \\
Q1231+2924  & 1.943 & C~IV							 & \\
				& 1.167 & C~IV, Al~III, Fe~II     & \\
Q1247+2657  & 1.959 & H~I, C~IV, Si~IV        & \\
				& 1.408 & C~IV							 & \\
				& 1.223 & Fe~II						 & \\
				& 0.762 & Mg~II						 & \\
Q1326+3923  & 2.150 & H~I, C~IV, Si~IV        & \\
				& 2.131 & H~I, Si~II, C~IV, Al~II, Si~II, Si~IV, Fe~II & \\
				& 2.088 & H~I                     & \tablenotemark{C} \\
Q1329+4117  & 1.940 & H~I, N~V, C~IV          & \tablenotemark{A} \\
				& 1.600 & H~I, C~IV               & \\
				& 1.471 & C~IV							 & \\
				& 0.922 & Mg~II						 & \\
Q1331+1704  & 1.776 & H~I, C~II, C~IV, O~I, Si~II, Si~IV & \tablenotemark{B} \\
			   & 0.745 & Mg~II, Fe~II				 & \\
Q1418+2254  & 2.191 & H~I, C~IV, Si~IV			 & \tablenotemark{A} \\
			   & 1.874 & H~I, C~IV, Si~IV			 & \\
Q1422+4224  & 1.951 & H~I                     & \\
Q1435+6349  & 1.923 & H~I, C~II, C~IV, Al~II, Si~II, Si~IV & \\
Q1517+2356  & 1.416 & C~IV							 & \\
Q1542+3104  & 2.076 & H~I, C~II, C~IV, Al~II, Al~III, Si~IV & \\ 
				& 1.725 & H~I, C~IV               & \\
Q1542+5408\tablenotemark{D} & 2.224 & H~I, C~IV, N~V, O~VI    & \tablenotemark{E} \\
Q1559+0853  & 2.281 & H~I, C~IV, N~V, Si~IV	 & \tablenotemark{A} \\
				& 2.252 & H~I, C~II, C~IV, Si~II  & \\
				& 2.122 & H~I, C~IV, Si~IV        & \\
            & 1.842 & H~I, C~II, C~IV, Al~II, Si~II  & \tablenotemark{B} \\
Q1618+5303  & 2.125 & H~I, C~II, Al,~II, Si~II	& \\
	         & 2.109 & H~I, C~II, C~IV, Al~II, Si~II, Si~IV & \\
Q1626+6433  & 2.292 & H~I, C~IV, Si~IV        & \\ 
				& 2.245 & H~I, C~IV					 & \\
				& 2.110 & H~I, C~II, C~IV, Al~II, Al~III, Si~IV & \\
				& 2.099 & C~IV							 & \\
				& 2.055 & H~I, C~IV, Si~IV			 & \\
				& 1.927 & H~I, C~IV, Al~II, Al~III, Si~II, Si~IV & \\
Q1632+3209  & 2.350 & H~I, C~IV, Si~IV        & \tablenotemark{A} \\
				& 2.258 & C~IV							 & \\
			   & 2.092 & H~I, C~II, C~IV, Al~II, Al~III, 
                      Si~II & \tablenotemark{B} \\
Q1649+4007  & 1.891 & H~I, C~IV, Al~II, Al~III, Si~II, Si~IV & \\
				& 1.799 & H~I, C~IV					 & \\
				& 1.638 & C~IV, Al~III				 & \\
				& 0.499 & Mg~II, Fe~II            & \\
Q1703+5350  & 2.370 & H~I, C~IV, N~V, Si~IV   & \tablenotemark{A} \\
				& 2.338 & C~IV, N~V, Si~IV			 & \\
				& 2.300 & C~IV, Si~IV				 & \\
				& 1.017 & Mg~II, Fe~II				 & \\
				& 0.699 & Mg~II, Fe~II				 & \\
				& 0.679 & Mg~II, Fe~II				 & \\
Q1705+7101  & 0.713 & Mg~II, Fe~II				 & \\
Q1716+4619  & 2.132 & H~I, C~IV, N~V, Si~IV   & \tablenotemark{A} \\ 
				& 2.049 & H~I, C~IV					 & \\
				& 1.563 & C~IV							 & \\
				& 0.980 & Mg~II						 & \\
				& 0.878 & Mg~II						 & \\
Q1720+2501  & 0.925 & Mg~II						 & \\
				& 0.817 & Mg~II						 & \\
Q1754+3818  & 1.784 & H~I, C~II, C~IV, O~I, Al~II, Al~III, 
                      Si~II, Si~IV & \tablenotemark{B} \\
Q1833+5811  & 2.038 & H~I, C~IV               & \tablenotemark{A} \\
Q1834+6117  & 2.210 & H~I, C~IV					 & \\
				& 1.100 & Mg~II, Fe~II				 & \\
Q2044--1650 & 1.920 & H~I, C~IV, Si~IV        & \\
Q2103+1843  & 2.110 & H~I, C~IV, Si~IV			 & \\
				& 1.964 & H~I, C~IV, Si~IV			 & \\
Q2134+0028	& 0.629 & Mg~II, Fe~II				 & \\
Q2134+1531  & 2.045 & C~IV							 & \\
				& 1.475 & C~IV							 & \\
				& 1.181 & Fe~II						 & \\
Q2140+2402  & 2.163 & H~I, C~II, C~IV, O~I, Al~II, Si~II & \tablenotemark{A} \\
				& 0.962 & Mg~II						 & \\
Q2150+0522  & 1.990 & H~I, C~IV, N~V			 & \tablenotemark{A} \\
				& 1.883 & H~I, C~II, O~I, Al~II, Si~II   & \\
				& 1.730 & H~I, C~IV					 & \\
Q2157--0036 & 1.963 & H~I, C~IV, N~V			 & \tablenotemark{A} \\
				& 1.650 & C~IV							 & \\
Q2241--2418 & 0.754 & Mg~II, Fe~II				 & \\
Q2245+2531  & 1.991 & H~I, C~IV				    & \\
				& 0.422 & Mg~II                   & \\
Q2310+0018\tablenotemark{D}  & 2.048 & H~I, C~IV, Si~IV   & \\
				& 1.999 & C~IV							 & \\
				& 1.971 & C~IV, N~V, Si~IV			 & \tablenotemark{E} \\
				& 1.902 & H~I, C~IV					 & \\
Q2332+2917  & 2.066 & H~I, C~IV, N~V			 & \tablenotemark{A} \\
				& 1.874 & H~I, C~IV					 & \\
				& 1.737 & C~IV, Si~IV				 & \\
				& 1.671 & C~IV							 & \\
				& 0.968 & Mg~II, Fe~II				 & \\
\enddata
\tablenotetext{A}{\footnotesize{Likely associated with QSO}}
\tablenotetext{B}{\footnotesize{Damped Lyman Alpha system}}
\tablenotetext{C}{\footnotesize{The strong H~I absorption at this redshift
shows no apparent metal absorption, and may instead be a blend}}
\tablenotetext{D}{\footnotesize{BAL QSO}}
\tablenotetext{E}{\footnotesize{BAL absorption}}

\end{deluxetable}

\clearpage
\begin{deluxetable}{lcccccccccccccc}
\tablewidth{0pt}
\tablecaption{\label{emltab}Observed Quasar Emission Line Peak Wavelengths}
\tabletypesize{\scriptsize}
\rotate
\tablehead{
\colhead{Name} &
\colhead{N~II} &
\colhead{Fe~III} &
\colhead{C~III$^{*}$} & 
\colhead{\lya} &
\colhead{N~V} &
\colhead{Si~II} &
\colhead{O~I} &
\colhead{C~II} &
\colhead{Si~IV} &
\colhead{C~IV} &
\colhead{He~II} &
\colhead{O~III$]$} &
\colhead{Al~III} &
\colhead{C~III$]$} 
\\
\colhead{(1950)} &
\colhead{He~II} &
\colhead{} &
\colhead{} & 
\colhead{} &
\colhead{} &
\colhead{} &
\colhead{Si~II} &
\colhead{} &
\colhead{O~IV$]$} &
\colhead{} &
\colhead{} &
\colhead{} &
\colhead{} &
\colhead{} 
\\
\colhead{ } &
\colhead{$\sim$1072} &
\colhead{1123} &
\colhead{1176} & 
\colhead{1216} &
\colhead{1240} &
\colhead{1263} &
\colhead{1306} &
\colhead{1335} &
\colhead{1400} &
\colhead{1549} &
\colhead{1640} &
\colhead{1664} &
\colhead{1860} &
\colhead{1909} 
}
\startdata
Q0001--2340 & 3538.38 & \nodata & \nodata & 3967.11 & \nodata & \nodata & 4260.14 & 4356.04 & 4540.78 & 5053.71 & \nodata & \nodata & \nodata & \nodata \\
Q0014--0420 & \nodata & \nodata & \nodata & 3592.63 & 3659.56 & \nodata & 3865.82 & 3954.43 & 4137.73 & 4576.02 & \nodata & 5558.63 & 5503.24 & 5649.53 \\
Q0049+0124 & \nodata & \nodata & \nodata & 4008.89 & \nodata & \nodata & 4305.66 & \nodata & 4612.29 & 5103.42 & \nodata & \nodata & \nodata & \nodata \\
Q0109+0213 & 3553.20 & 3735.64 & \nodata & 4069.83 & 4151.09 & \nodata & 4365.39 & \nodata & 4672.75 & 5188.09 & 5496.71 & \nodata & \nodata & \nodata \\
Q0150-2015 & \nodata & 3511.28 & \nodata & 3820.70 & \nodata & \nodata & 4102.96 & 4193.47 & 4399.05 & 4864.86 & 5146.26 & 5236.25 & 5838.29 & \nodata \\
Q0153+7428 & \nodata & \nodata & \nodata & 4059.21 & 4135.19 & \nodata & 4355.94 & 4459.03 & 4667.56 & 5166.69 & 5472.18 & 5577.23 & \nodata & \nodata \\
Q0218+3707 & 3647.41 & \nodata & \nodata & 4154.20 & \nodata & 4318.01 & 4461.55 & 4554.49 & 4775.75 & 5289.66 & 5595.30 & 5670.37 & \nodata & \nodata \\
Q0226--0350 & \nodata & 3453.79 & \nodata & 3721.64 & \nodata & \nodata & 4017.07 & 4107.23 & 4288.43 & 4740.79 & 5026.28 & 5090.13 & 5693.01 & 5846.98 \\
Q0248+3402 & 3425.52 & 3612.57 & \nodata & 3919.65 & \nodata & 4061.37 & 4212.22 & 4306.73 & 4512.03 & 4992.58 & \nodata & \nodata & \nodata & \nodata \\
Q0348+0610 & \nodata & \nodata & \nodata & 3715.06 & \nodata & \nodata & 3999.62 & \nodata & 4284.32 & 4725.99 & 4996.42 & \nodata & 5676.31 & 4847.23 \\
Q0421+0157 & \nodata & \nodata & \nodata & 3702.54 & \nodata & \nodata & \nodata & \nodata & 4267.17 & 4725.57 & \nodata & \nodata & 5661.76 & 5826.89 \\
Q0424--1309 & \nodata & \nodata & \nodata & 3845.37 & \nodata & \nodata & \nodata & \nodata & \nodata & 4898.42 & \nodata & \nodata & \nodata & \nodata \\
Q0450--1310 & 3473.08 & \nodata & \nodata & 3952.42 & \nodata & \nodata & \nodata & \nodata & 4537.24 & 5031.81 & 5342.14 & \nodata & \nodata & \nodata \\
Q0726+2531 & \nodata & \nodata & 3880.68 & 4011.13 & \nodata & \nodata & \nodata & \nodata & 4617.62 & 5107.65 & \nodata & \nodata & \nodata & \nodata \\
Q0743+6601 & 3423.58 & 3581.04 & \nodata & 3892.27 & \nodata & \nodata & 4176.63 & 4280.26 & 4472.10 & 4921.70 & \nodata & \nodata & \nodata & \nodata \\
Q0748+6105 & 3718.48 & \nodata & \nodata & 4239.84 & 4319.47 & 4404.59 & 4547.94 & 4653.01 & 4872.52 & 5390.25 & \nodata & \nodata & \nodata & \nodata \\
Q0752+3429 & \nodata & \nodata & \nodata & 3791.02 & 3871.45 & \nodata & 4073.31 & 4166.24 & 4364.53 & 4835.17 & \nodata & \nodata & 5806.45 & \nodata \\
Q0800+3031 & \nodata & \nodata & 3553.25 & 3676.34 & 3746.62 & \nodata & 3946.64 & \nodata & 4231.40 & 4686.19 & 4946.86 & 5036.85 & \nodata & 5781.41 \\
Q0836+7104 & \nodata & 3571.99 & \nodata & 3865.05 & \nodata & \nodata & 4154.83 & \nodata & 4453.38 & 4927.89 & \nodata & \nodata & \nodata & \nodata \\
Q0854+3324 & \nodata & \nodata & \nodata & 4055.97 & 4136.37 & 4218.57 & 4361.06 & 4459.36 & 4672.29 & 5174.13 & 5477.93 & 5558.73 & \nodata & \nodata \\
Q0907+3811 & \nodata & \nodata & \nodata & 3837.07 & \nodata & \nodata & \nodata & 4215.64 & 4417.31 & 4892.09 & \nodata & \nodata & \nodata & \nodata \\
Q0936+3653 & \nodata & \nodata & \nodata & 3670.74 & \nodata & \nodata & 3946.93 & \nodata & 4229.70 & 4680.11 & \nodata & \nodata & 5620.24 & 5768.60 \\
Q0937--1818 & \nodata & \nodata & \nodata & 4091.86 & 4172.38 & \nodata & 4488.98 & \nodata & 4707.92 & 5211.50 & \nodata & \nodata & \nodata & \nodata \\
Q1023+3009 & 3584.52 & 3739.53 & \nodata & 4058.52 & 4123.40 & 4202.68 & 4348.69 & \nodata & 4661.56 & 5150.80 & \nodata & \nodata & \nodata & \nodata \\
Q1103+6416 & 3422.10 & \nodata & \nodata & 3895.55 & \nodata & \nodata & 4195.05 & 4291.91 & 4462.29 & 4936.25 & \nodata & \nodata & \nodata & \nodata \\
Q1116+2106 & \nodata & \nodata & \nodata & 4207.48 & \nodata & \nodata & \nodata & \nodata & 4823.21 & 5349.04 & \nodata & \nodata & \nodata & \nodata \\
Q1122--1648 & 3619.81 & 3827.42 & 3999.26 & 4129.90 & 4212.21 & \nodata & \nodata & \nodata & 4754.99 & 5259.91 & \nodata & \nodata & \nodata & \nodata \\
Q1130+3135 & \nodata & \nodata & 3876.87 & 4009.12 & 4089.53 & \nodata & 4307.60 & \nodata & 4615.85 & 5107.89 & 5406.98 & \nodata & \nodata & \nodata \\
Q1147+6556 & \nodata & \nodata & \nodata & 3908.56 & \nodata & \nodata & \nodata & \nodata & \nodata & 4976.16 & \nodata & \nodata & \nodata & \nodata \\
Q1222+2251 & 3278.00 & \nodata & \nodata & 3703.63 & 3773.37 & \nodata & 3988.37\tablenotemark{A} & 4072.34 & 4270.70 & 4717.24 & \nodata & \nodata & 5665.21 & 5815.42 \\
Q1224--0812 & \nodata & \nodata & \nodata & 3842.52 & \nodata & \nodata & \nodata & \nodata & \nodata & 4853.82 & \nodata & \nodata & \nodata & \nodata \\
Q1224+2905 & \nodata & \nodata & \nodata & 3958.05 & \nodata & \nodata & \nodata & \nodata & 4550.13 & 5029.28 & \nodata & \nodata & \nodata & \nodata \\
Q1225+3145 & \nodata & \nodata & \nodata & 3868.93 & \nodata & \nodata & \nodata & \nodata & \nodata & 4918.67 & \nodata & \nodata & \nodata & \nodata \\
Q1231+2924 & \nodata & \nodata & \nodata & 3665.14 & \nodata & 3806.59 & 3938.70 & 4024.38 & 4220.80 & 4668.24 & \nodata & \nodata & \nodata & 5752.26 \\
Q1247+3145 & 3269.99 & 3423.82 & \nodata & 3695.20 & 3769.19 & 3837.80 & 3972.72 & 4058.11 & 4251.21 & 4708.09 & \nodata & 5058.89 & 5657.24 & 5807.34 \\
Q1251+2636 & \nodata & 3407.63 & \nodata & 3685.96 & \nodata & \nodata & 3960.87 & \nodata & 4240.61 & 4689.39 & 4966.96 & \nodata & 5649.70 & 5807.29 \\
Q1307+4617 & 3352.25 & 3515.25 & \nodata & 3804.88 & \nodata & 3851.99 & 4089.20 & 4192.29 & 4369.52 & 4834.06 & \nodata & \nodata & \nodata & \nodata \\
Q1312+7837 & 3245.92 & 3372.81 & \nodata & 3648.05 & 3719.52 & 3796.36 & 3916.10 & \nodata & 4200.26 & 4651.81 & 4920.32 & \nodata & 5583.90 & 5734.26 \\
Q1326+3923 & 3551.49 & 3712.89 & \nodata & 4033.84 & 4112.78 & \nodata & \nodata & \nodata & 4642.02 & 5128.33 & \nodata & \nodata & \nodata & \nodata \\
Q1329+4117 & \nodata & \nodata & \nodata & 3562.93 & 3633.73 & \nodata & 3828.06 & 3911.57 & 4102.21 & 4543.41 & 4810.14 & 4875.95 & \nodata & 5601.05 \\
Q1331+1704 & \nodata & \nodata & \nodata & 3745.32 & 3810.23 & 3891.61 & 4036.10\tablenotemark{A} & \nodata & \nodata & 4767.11 & \nodata & \nodata & \nodata & \nodata \\
Q1416+0906 & \nodata & 3378.23 & \nodata & 3657.23 & 3721.61 & 3798.49 & \nodata & \nodata & 4208.03 & 4657.94 & \nodata & \nodata & \nodata & 5739.54 \\
Q1418+2254 & 3396.61 & \nodata & \nodata & 3870.70 & \nodata & \nodata & 4158.90 & 4252.89 & 4455.25 & 4927.64 & \nodata & 5297.17 & \nodata & \nodata \\
Q1422+4224 & \nodata & \nodata & \nodata & 3904.79 & \nodata & \nodata & \nodata & \nodata & 4491.29 & 4970.92 & \nodata & \nodata & \nodata & \nodata \\
Q1425--1338 & \nodata & \nodata & \nodata & 3672.73 & 3735.46 & \nodata & \nodata & \nodata & 4221.27 & 4665.66 & \nodata & \nodata & \nodata & 5751.41 \\
Q1435+6349 & 3257.09 & 3439.33 & 3598.14 & 3722.15 & \nodata & \nodata & 3995.12 & \nodata & 4285.63 & 4746.52 & \nodata & \nodata & \nodata & 5847.43 \\
Q1517+2356 & \nodata & \nodata & \nodata & 3527.90 & 3593.26 & \nodata & 3791.43 & \nodata & 4058.99 & 4492.05 & \nodata & \nodata & \nodata & 5533.73 \\
Q1542+3104 & \nodata & \nodata & 3851.17 & 3996.24 & \nodata & \nodata & \nodata & \nodata & 4585.67 & 5072.14 & \nodata & \nodata & \nodata & \nodata \\
Q1542+5408\tablenotemark{B} & 3601.08 & \nodata & \nodata & 4093.20 & \nodata & \nodata & 4400.36 & \nodata & 4707.39 & \nodata & \nodata & \nodata & \nodata & \nodata \\
Q1559+0853 & \nodata & \nodata & \nodata & 3974.32 & 4052.14 & \nodata & 4267.77 & \nodata & 4572.26 & 5054.89 & \nodata & \nodata & \nodata & \nodata \\
Q1611+4719 & \nodata & \nodata & \nodata & 4122.07 & \nodata & \nodata & 4406.98 & \nodata & 4693.79 & 5189.04 & \nodata & \nodata & \nodata & \nodata \\
Q1618+5303 & 3569.20 & 3760.92 & \nodata & 4072.03 & \nodata & \nodata & 4374.11 & \nodata & 4686.69 & 5182.64 & \nodata & 5579.00 & \nodata & \nodata \\
Q1626+6433 & 3531.17 & 3715.87 & \nodata & 4022.81 & 4091.56 & 4176.64 & 4324.98\tablenotemark{A} & 4419.10 & 4637.36 & 5123.94 & \nodata & \nodata & \nodata & \nodata \\
Q1632+3209 & 3578.04 & \nodata & \nodata & 4067.78 & 4149.19 & \nodata & 4370.22 & \nodata & 4678.08 & 5173.28 & \nodata & \nodata & \nodata & \nodata \\
Q1649+4007 & \nodata & \nodata & 3919.90 & 4057.01 & 4136.65 & \nodata & 4357.44 & \nodata & 4676.83 & 5171.99 & 5470.39 & 5553.93 & \nodata & \nodata \\
Q1703+5350 & \nodata & \nodata & \nodata & 4093.11 & 4173.13 & 4250.98 & 4395.64 & 4502.46 & 4716.50 & 5214.11 & 5512.69 & 5604.67 & \nodata & \nodata \\
Q1705+7101 & \nodata & \nodata & \nodata & 3659.56 & 3730.10 & \nodata & \nodata & \nodata & 4211.27 & 4661.54 & \nodata & \nodata & \nodata & 5745.21 \\
Q1716+4519 & \nodata & 3488.73 & \nodata & 3778.86 & \nodata & \nodata & 4066.54\tablenotemark{A} & 4150.57 & 4348.35 & 4791.79 & \nodata & \nodata & \nodata & \nodata \\
Q1720+2501 & 3478.00 & \nodata & \nodata & 3951.56 & 4025.86 & \nodata & \nodata & \nodata & 4543.90 & 5033.93 & 5332.45 & \nodata & \nodata & \nodata \\
Q1754+3818 & \nodata & 3547.88 & \nodata & 3834.49 & \nodata & \nodata & 4116.77 & 4210.43 & 4412.27 & 4884.23 & \nodata & \nodata & \nodata & \nodata \\
Q1833+5813 & 3251.34 & \nodata & \nodata & 3684.65 & \nodata & \nodata & 3965.69 & \nodata & 4232.23 & 4681.22 & \nodata & \nodata & \nodata & 5771.15 \\
Q1834+6117 & 3521.56 & 3672.31 & 3850.92 & 3981.05 & 4060.70 & 4136.95 & 4273.09 & \nodata & 4563.05 & 5075.61 & 5373.39 & \nodata & \nodata & \nodata \\
Q1848+6705 & \nodata & 3390.94 & 3555.86 & 3682.76 & 3746.20 & \nodata & 3947.47 & 4036.26 & 4228.47 & 4678.83 & \nodata & \nodata & \nodata & 5779.84 \\
Q2044--1650 & \nodata & 3295.56 & \nodata & 3572.47 & 3644.85 & \nodata & \nodata & \nodata & 4109.97 & 4552.26 & 4820.87 & \nodata & \nodata & 5607.79 \\
Q2103+1843 & \nodata & \nodata & \nodata & 3898.96 & 3978.71 & \nodata & 4187.53 & 4281.44 & 4488.13 & 4970.02 & 5259.21 & 5338.05 & \nodata & \nodata \\
Q2134+0028 & \nodata & \nodata & \nodata & 3574.56 & \nodata & \nodata & 3843.43 & 3921.12 & 4114.54 & 4548.77 & 4821.31 & \nodata & \nodata & 5610.11 \\
Q2134+1531 & \nodata & 3525.60 & 3681.23 & 3807.99 & 3882.19 & 3954.58 & 4090.31 & 4180.85 & 4383.55 & 4854.16 & 5137.08 & 5217.82 & \nodata & \nodata \\
Q2135+1326 & 3525.60 & 3705.06 & \nodata & 4007.78 & 4073.03 & 4160.02 & 4308.69 & 4408.41 & 4607.32 & 5092.70 & \nodata & \nodata & \nodata & \nodata \\
Q2140+2403 & \nodata & \nodata & \nodata & \nodata & \nodata & \nodata & 4132.89 & \nodata & 4426.43 & 4893.11 & \nodata & \nodata & \nodata & \nodata \\
Q2147--0825 & \nodata & \nodata & \nodata & 3796.25 & \nodata & \nodata & 4079.66 & \nodata & 4366.70 & 4836.31 & \nodata & \nodata & \nodata & \nodata \\
Q2150+0522 & \nodata & \nodata & \nodata & 3621.48 & 3694.10 & \nodata & \nodata & \nodata & 4169.89 & 4628.60 & 4887.75 & \nodata & 5539.35 & 5674.18 \\
Q2157-0036 & \nodata & \nodata & \nodata & 3601.79 & 3670.88 & \nodata & 3866.89 & \nodata & 4144.75 & 4586.64 & \nodata & 4929.17 & 5505.82 & 5654.17 \\
Q2241--2418 & \nodata & 3322.16 & \nodata & 3597.83 & 3666.74 & \nodata & 3866.26 & \nodata & 4143.69 & 4583.13 & \nodata & 4926.35 & \nodata & 5647.36 \\
Q2245+2531 & \nodata & 3547.02 & \nodata & 3837.61 & 3912.05 & \nodata & \nodata & \nodata & \nodata & 4892.12 & 5175.12 & 5250.30 & \nodata & \nodata \\
Q2310+0018\tablenotemark{B} & 3301.11 & \nodata & \nodata & 3745.31 & 3812.39 & 3892.14 & 4026.28 & \nodata & 4319.99 & 4769.51 & 5052.07 & \nodata & 5728.50  & \nodata \\
Q2310+3831 & 3412.83 & 3568.49 & \nodata & 3861.60 & 3935.84 & \nodata & 4151.16\tablenotemark{A} & \nodata & 4449.94 & 4923.74 & \nodata & 5286.97 & 5728.51 & \nodata \\
Q2320+0755 & 3309.54 & \nodata & \nodata & 3753.69 & 3824.23 & \nodata & 4035.43 & \nodata & 4322.25 & 4776.34 & 5057.92 & \nodata & \nodata & \nodata \\
Q2329--0204 & \nodata & \nodata & 3406.31 & 3520.29 & 3589.02 & \nodata & 3780.76 & \nodata & 4052.15 & 4482.69 & 4742.46 & 4812.01 & \nodata & 5523.808 \\
Q2332+2917 & 3267.11 & \nodata & \nodata & 3732.02 & \nodata & \nodata & 4017.15 & \nodata & 4294.97 & 4739.35 & \nodata & \nodata & \nodata & 5841.71 
\enddata
\tablenotetext{A}{The O~I--Si~II blend in this line is dominated by the Si~II
1309 transition}
\tablenotetext{B}{BAL QSO}
\end{deluxetable}

\begin{deluxetable}{lcccc}
\tablewidth{0pt}
\tablecaption{\label{emlztab}Effective Redshifts \& Calculated Rest Wavelengths}
\tablehead{
\colhead{Name} &
\colhead{$z_{eff}$} &
\colhead{$\lambda_{rest}$} &
\colhead{$\lambda_{rest}$} &
\colhead{$\lambda_{rest}$} 
\\
\colhead{(1950)} &
\colhead{} &
\colhead{(1071\AA)} &
\colhead{(1123\AA)} &
\colhead{(1176\AA)} 
}
\startdata
Q0001--2340 &  2.2587 &  1085.84 &  \nodata &  \nodata \\
Q0014--0420 &  1.9996 &  \nodata &  \nodata &  \nodata \\
Q0049+0124 &  2.2957 &  \nodata &  \nodata &  \nodata \\
Q0109+0213 &  2.3460 &  1061.94 &  1116.46 &  \nodata \\
Q0150--2015 &  2.1414 &  \nodata &  1117.74 &  \nodata \\
Q0153+7428 &  2.3383 &  \nodata &  \nodata &  \nodata \\
Q0218+3707 &  2.4136 &  1068.50 &  \nodata &  \nodata \\
Q0226--0350 &  2.0649 &  \nodata &  1126.89 &  \nodata \\
Q0248+3402 &  2.2227 &  1062.93 &  1120.97 &  \nodata \\
Q0348+0610 &  1.9809 &  \nodata &  \nodata &  \nodata \\
Q0421+0157 &  2.0480 &  \nodata &  \nodata &  \nodata \\
Q0424--1309 &  2.1623 &  \nodata &  \nodata &  \nodata \\
Q0450--1310 &  2.2493 &  1068.88 &  \nodata &  \nodata \\
Q0726+2531 &  2.2981 &  \nodata &  \nodata &  1176.64 \\
Q0743+6601 &  2.1954 &  1071.42 &  1120.70 &  \nodata \\
Q0748+6105 &  2.4836 &  1067.41 &  \nodata &  \nodata \\
Q0752+3429 &  2.1200 &  \nodata &  \nodata &  \nodata \\
Q0800+3031 &  2.0233 &  \nodata &  \nodata &  1175.30 \\
Q0836+7104 &  2.1805 &  \nodata &  1123.08 &  \nodata \\
Q0854+3324 &  2.3388 &  \nodata &  \nodata &  \nodata \\
Q0907+3811 &  2.1567 &  \nodata &  \nodata &  \nodata \\
Q0936+3653 &  2.0211 &  \nodata &  \nodata &  \nodata \\
Q0937--1818 &  2.3789 &  \nodata &  \nodata &  \nodata \\
Q1023+3009 &  2.3292 &  1076.69 &  1123.25 &  \nodata \\
Q1103+6416 &  2.2009 &  1069.09 &  \nodata &  \nodata \\
Q1116+2106 &  2.4528 &  \nodata &  \nodata &  \nodata \\
Q1122--1648 &  2.3963 &  1065.80 &  1126.93 &  1177.52 \\
Q1130+3135 &  2.2975 &  \nodata &  \nodata &  1175.71 \\
Q1147+6556 &  2.2134 &  \nodata &  \nodata &  \nodata \\
Q1222+2251 &  2.0476 &  1075.59 &  \nodata &  \nodata \\
Q1224--0812 &  2.1467 &  \nodata &  \nodata &  \nodata \\
Q1224+2905 &  2.2506 &  \nodata &  \nodata &  \nodata \\
Q1225+3145 &  2.1785 &  \nodata &  \nodata &  \nodata \\
Q1231+2924 &  2.0143 &  \nodata &  \nodata &  \nodata \\
Q1247+3145 &  2.0399 &  1075.70 &  1126.31 &  \nodata \\
Q1251+2636 &  2.0327 &  \nodata &  1123.65 &  \nodata \\
Q1307+4617 &  2.1154 &  1076.04 &  1128.36 &  \nodata \\
Q1312+7837 &  2.0015 &  1081.44 &  1123.71 &  \nodata \\
Q1326+3923 &  2.3151 &  1071.30 &  1119.98 &  \nodata \\
Q1329+4117 &  1.9314 &  \nodata &  \nodata &  \nodata \\
Q1331+1704 &  2.0804 &  \nodata &  \nodata &  \nodata \\
Q1416+0906 &  2.0060 &  \nodata &  1123.84 &  \nodata \\
Q1418+2254 &  2.1834 &  1066.99 &  \nodata &  \nodata \\
Q1422+4224 &  2.2095 &  \nodata &  \nodata &  \nodata \\
Q1425--1338 &  2.0146 &  \nodata &  \nodata &  \nodata \\
Q1435+6349 &  2.0617 &  1063.82 &  1123.34 &  1175.21 \\
Q1517+2356 &  1.9000 &  \nodata &  \nodata &  \nodata \\
Q1542+3104 &  2.2788 &  \nodata &  \nodata &  1174.58 \\
Q1542+5408\tablenotemark{A} &  2.3660 &  1069.85 &  \nodata &  \nodata \\
Q1559+0853 &  2.2667 &  \nodata &  \nodata &  \nodata \\
Q1611+4719 &  2.3667 &  \nodata &  \nodata &  \nodata \\
Q1618+5303 &  2.3488 &  1065.81 &  1123.05 &  \nodata \\
Q1626+6433 &  2.3081 &  1067.42 &  1123.25 &  \nodata \\
Q1632+3209 &  2.3438 &  1070.06 &  \nodata &  \nodata \\
Q1649+4007 &  2.3374 &  \nodata &  \nodata &  1174.54 \\
Q1703+5350 &  2.3667 &  \nodata &  \nodata &  \nodata \\
Q1705+7101 &  2.0089 &  \nodata &  \nodata &  \nodata \\
Q1716+4519 &  2.1060 &  \nodata &  1123.24 &  \nodata \\
Q1720+2501 &  2.2486 &  1070.60 &  \nodata &  \nodata \\
Q1754+3818 &  2.1528 &  \nodata &  1125.30 &  \nodata \\
Q1833+5813 &  2.0270 &  1074.12 &  \nodata &  \nodata \\
Q1834+6117 &  2.2726 &  1076.06 &  1122.12 &  1176.70 \\
Q1848+6705 &  2.0235 &  \nodata &  1121.54 &  1176.09 \\
Q2044--1650 &  1.9382 &  \nodata &  1121.64 &  \nodata \\
Q2103+1843 &  2.2072 &  \nodata &  \nodata &  \nodata \\
Q2134+0028 &  1.9391 &  \nodata &  \nodata &  \nodata \\
Q2134+1531 &  2.1322 &  \nodata &  1125.59 &  1175.28 \\
Q2135+1326 &  2.2935 &  1070.48 &  1124.97 &  \nodata \\
Q2140+2403 &  2.1617 &  \nodata &  \nodata &  \nodata \\
Q2147--0825 &  2.1217 &  \nodata &  \nodata &  \nodata \\
Q2150+0522 &  1.9792 &  \nodata &  \nodata &  \nodata \\
Q2157--0036 &  1.9611 &  \nodata &  \nodata &  \nodata \\
Q2241--2418 &  1.9591 &  \nodata &  1122.70 &  \nodata \\
Q2245+2531 &  2.1560 &  \nodata &  1123.91 &  \nodata \\
Q2310+0018\tablenotemark{A} &  2.0805 &  1071.60 &  \nodata &  \nodata \\
Q2310+3831 &  2.1771 &  1074.19 &  1123.18 &  \nodata \\
Q2320+0755 &  2.0860 &  1072.45 &  \nodata &  \nodata \\
Q2329--0204 &  1.8937 &  \nodata &  \nodata &  1177.14 \\
Q2332+2917 &  2.0665 &  1065.42 &  \nodata &  \nodata \\
\tableline
$N$ & 79 & 29 & 27 & 11  \\
Mean $\lambda_{rest}$ & 2.1661 & 1070.95 & 1123.17 & 1175.88 \\
$\sigma$ &  & 5.41 & 2.65 & 1.01 \\
$\sigma / \sqrt{N}$ &  & 1.00 & 0.51 & 0.30 
\enddata
\tablenotetext{A}{BAL QSO}
\end{deluxetable} 

\begin{deluxetable}{lccc}
\tablewidth{0pt}
\tablecaption{\label{emlextras}Additional Emission Lines}
\tablehead{
\colhead{Name} &
\colhead{$\lambda_{obs}$} &
\colhead{$\lambda_{rest}$\tablenotemark{A}} &
\colhead{Transition} 
\\
\colhead{(1950)} &
\colhead{(\AA)} &
\colhead{(\AA)} &
\colhead{} 
}
\startdata
Q1023+3009  &  3978  & 1195 & Si~II \\
Q1222+2251  &  4478  & 1469 & \nodata  \\
Q1225+3145  &  4196  & 1320 & \nodata  \\
Q1307+4617  &  4593  & 1474 & \nodata  \\
Q1312+7837  &  5253  & 1750 & N~III$]$ \\
Q1425--1338 &  3431  & 1138 & \nodata  \\
Q1517+2556  &  4663  & 1608 & Fe~II \\
Q1559+0853  &  3691  & 1130 & \nodata  \\
Q1611+4719  &  4407  & 1309 & Si~II  \\
Q1705+7101  &  3364  & 1118 & \nodata  \\
Q1716+4519  &  5361  & 1726 & \nodata  \\
Q2044--1650 &  3367  & 1146 & \nodata  \\
Q2134+0028  &  3480  & 1184 & \nodata  \\
Q2134+1531  &  5607  & 1790 & \nodata  \\
Q2140+2403  &  3570  & 1129 & \nodata  \\
Q2245+2531  &  3667  & 1162 & \nodata  \\
Q2310+0018\tablenotemark{B} & 5514  & 1790 & \nodata \\
\enddata
\tablenotetext{A}{Obtained using $z_{eff}$ from Table \ref{emlztab}}
\tablenotetext{B}{BAL QSO}
\end{deluxetable}

\begin{figure}
\epsscale{0.8}
\plotone{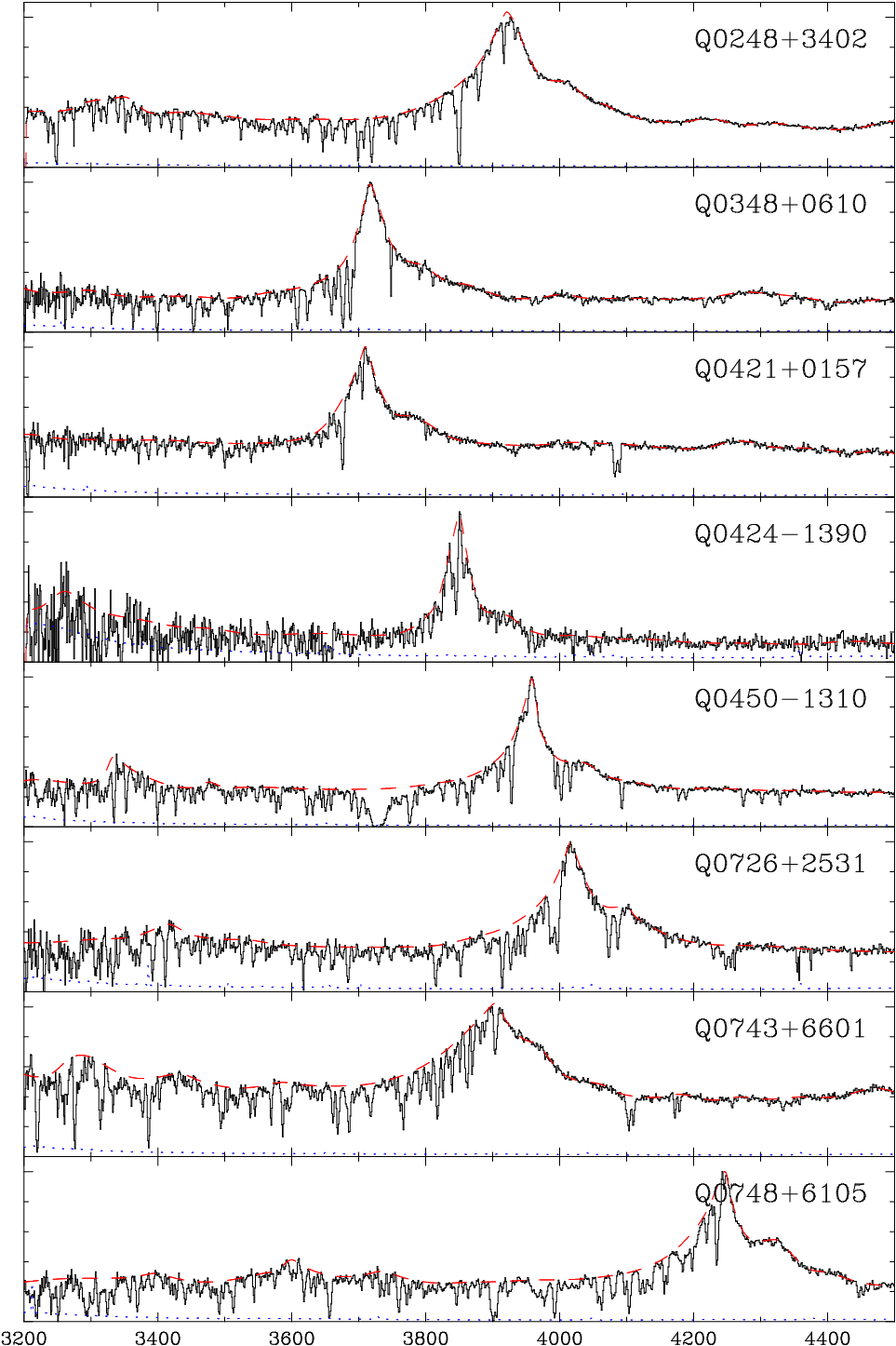}
\caption{Same as for Figure \ref{specfig}.}
\end{figure}

\begin{figure}
\epsscale{0.8}
\plotone{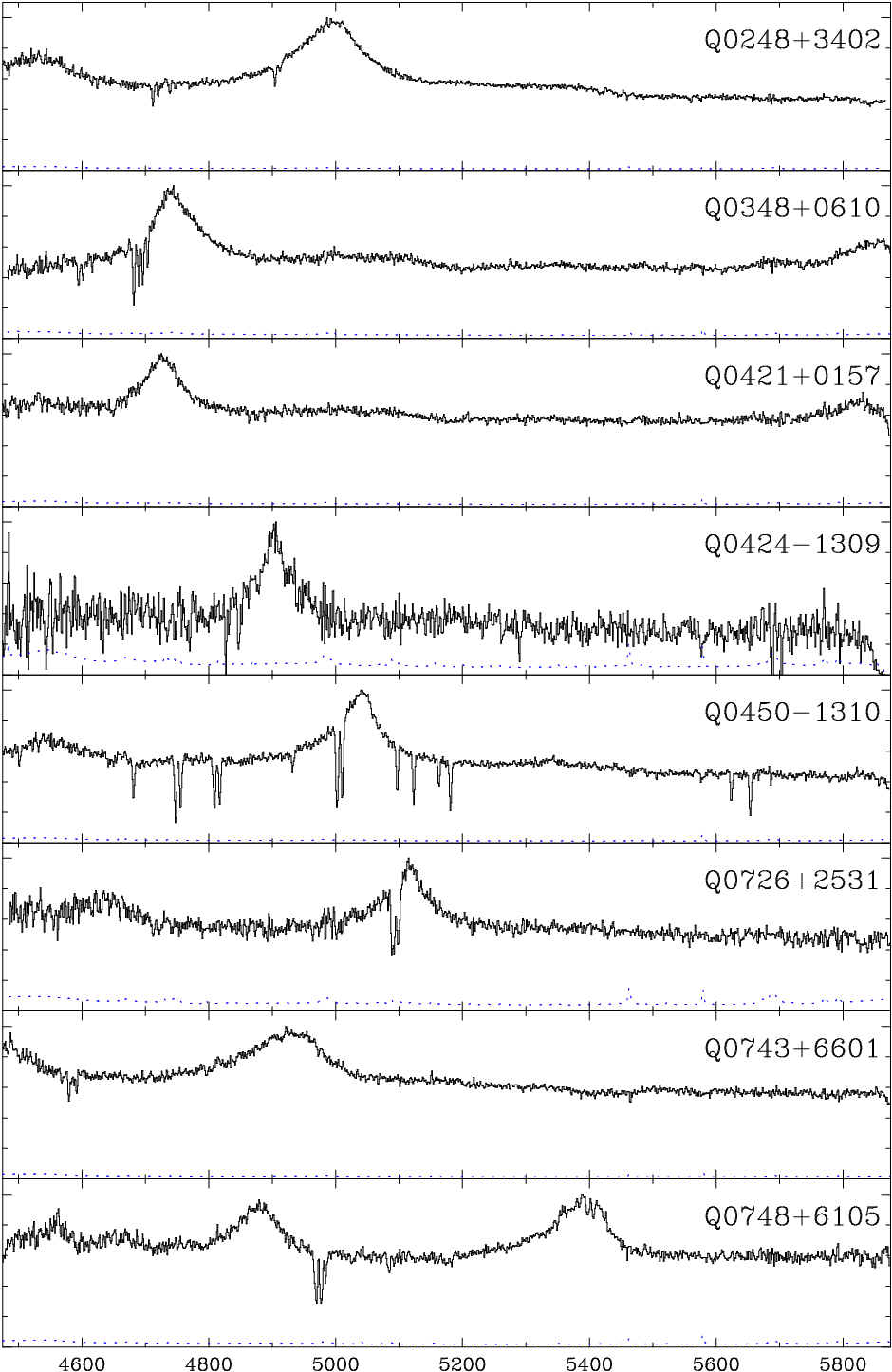}
\caption{Same as for Figure \ref{redspecfig}.}
\end{figure}

\clearpage

\begin{figure}
\epsscale{0.8}
\plotone{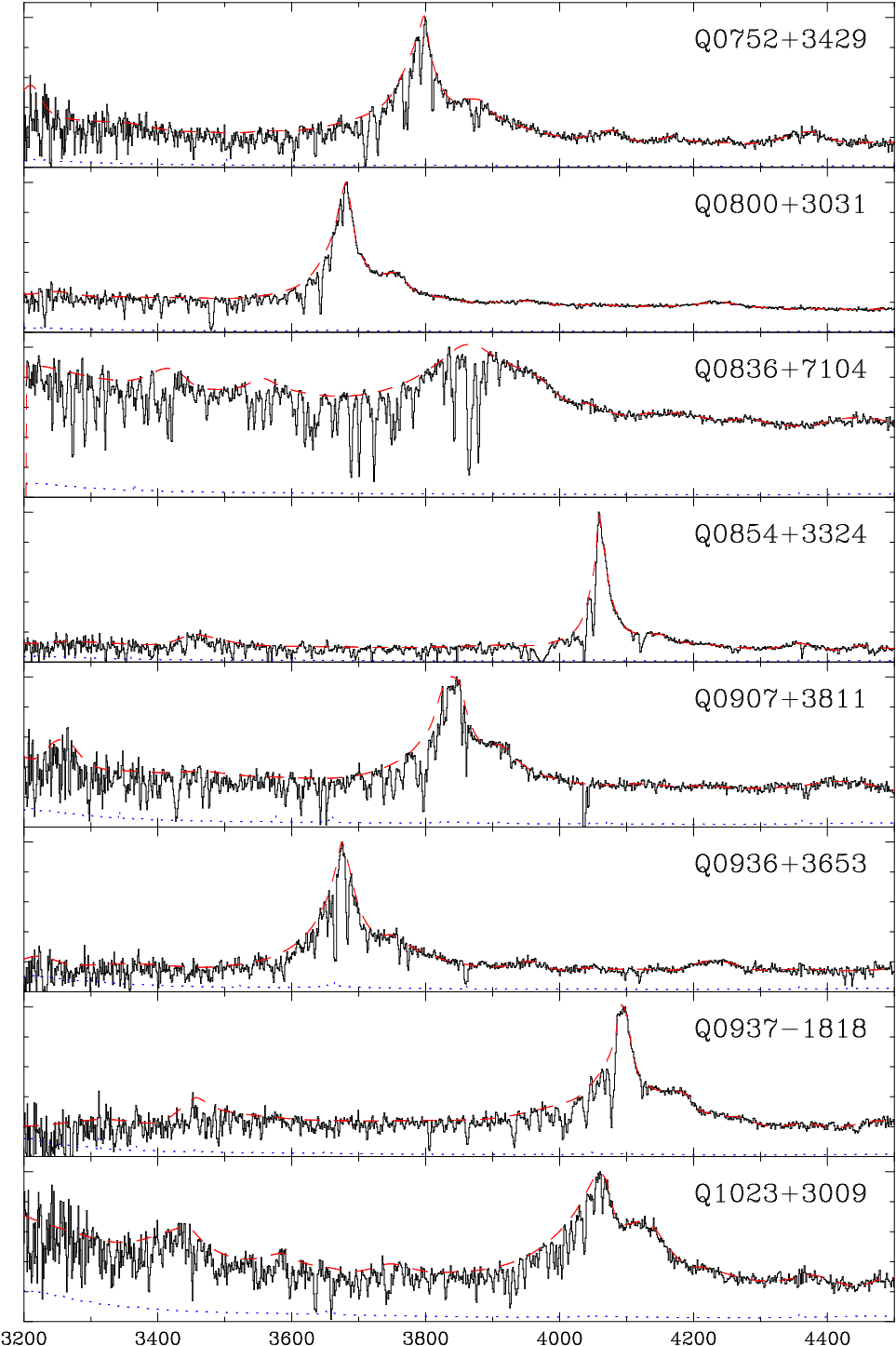}
\caption{Same as for Figure \ref{specfig}.}
\end{figure}

\begin{figure}
\epsscale{0.8}
\plotone{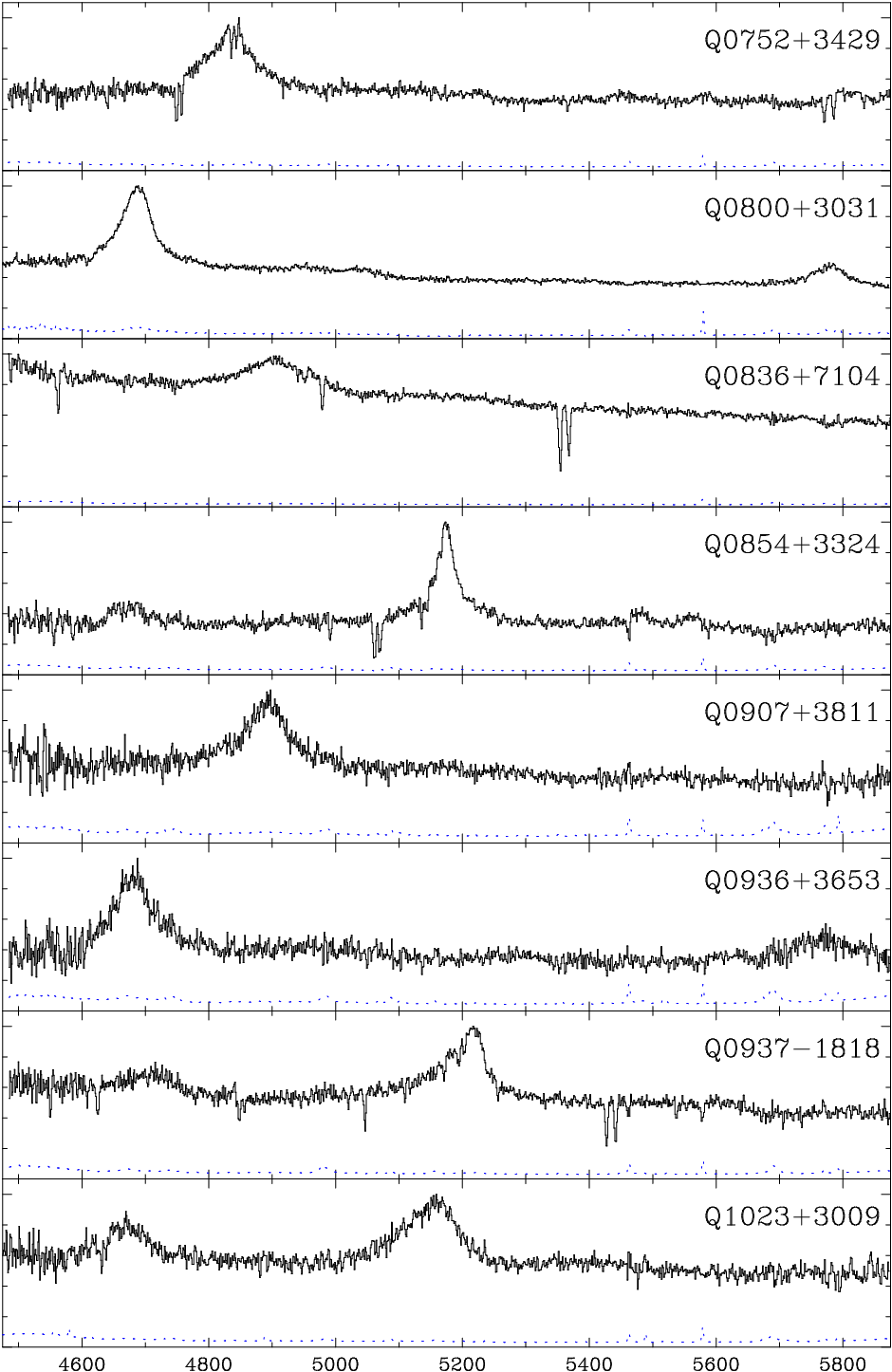}
\caption{Same as for Figure \ref{redspecfig}.}
\end{figure}

\clearpage

\begin{figure}
\epsscale{0.8}
\plotone{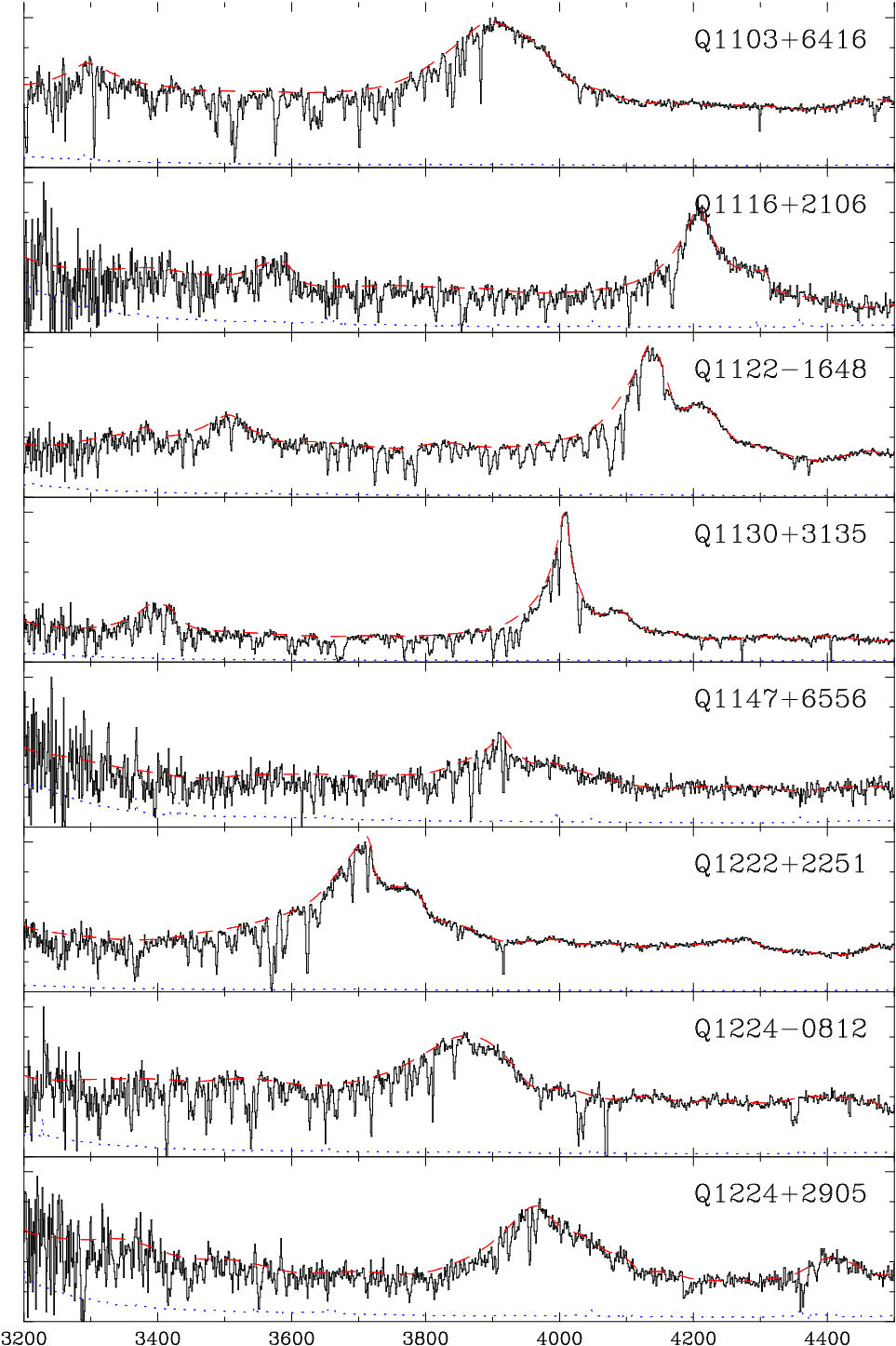}
\caption{Same as for Figure \ref{specfig}.}
\end{figure}

\begin{figure}
\epsscale{0.8}
\plotone{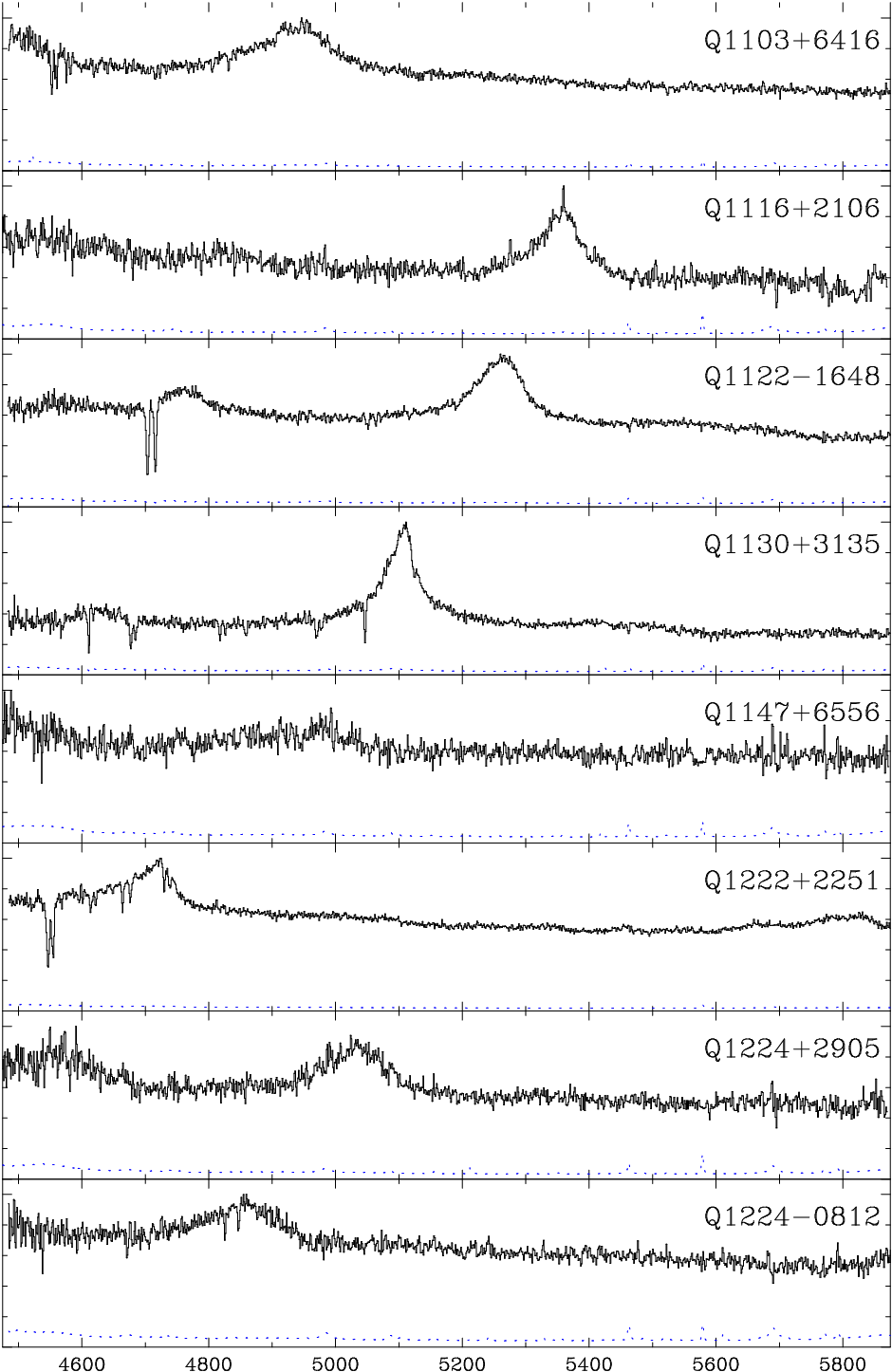}
\caption{Same as for Figure \ref{redspecfig}.}
\end{figure}

\clearpage

\begin{figure}
\epsscale{0.8}
\plotone{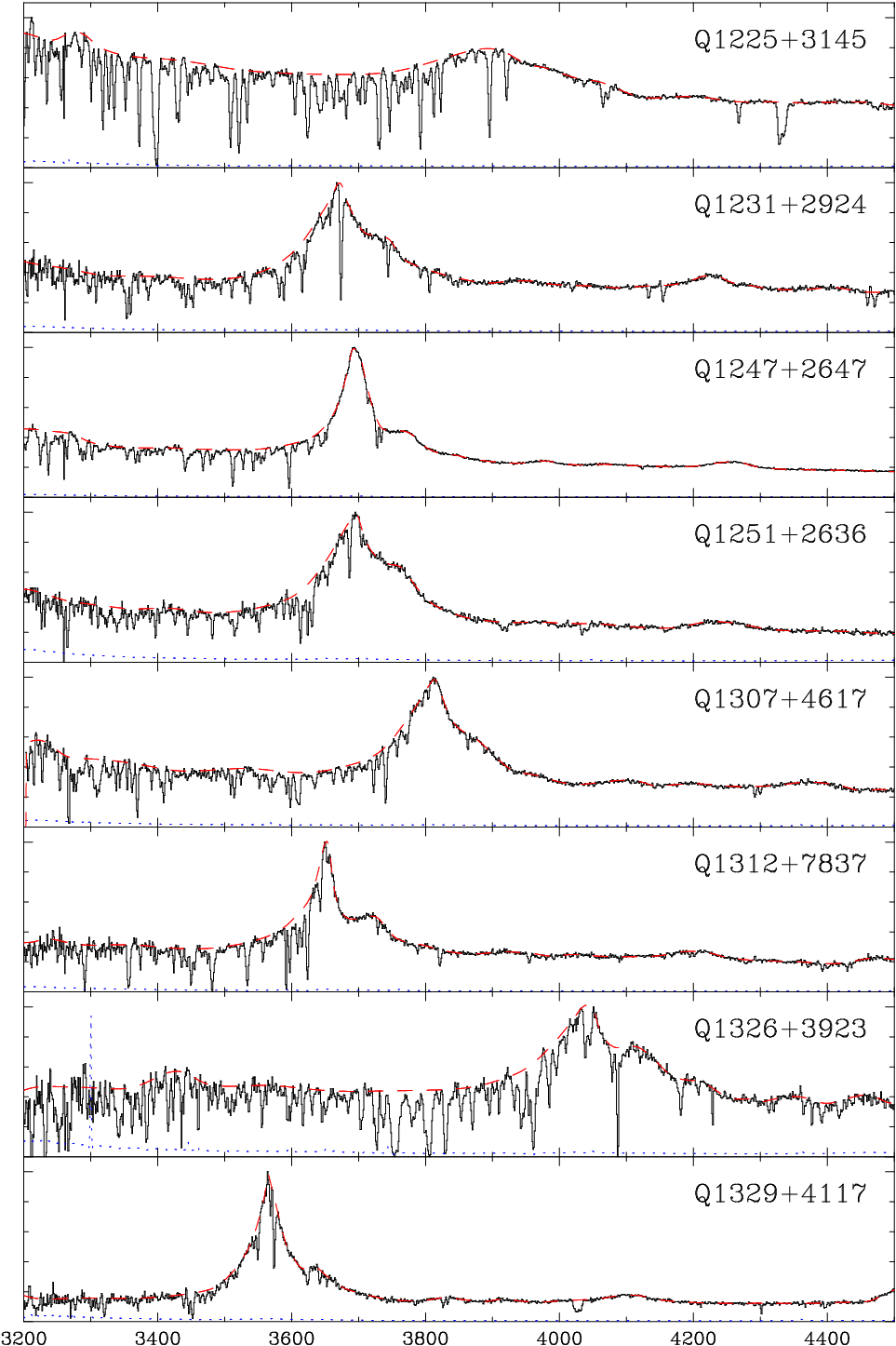}
\caption{Same as for Figure \ref{specfig}.}
\end{figure}

\begin{figure}
\epsscale{0.8}
\plotone{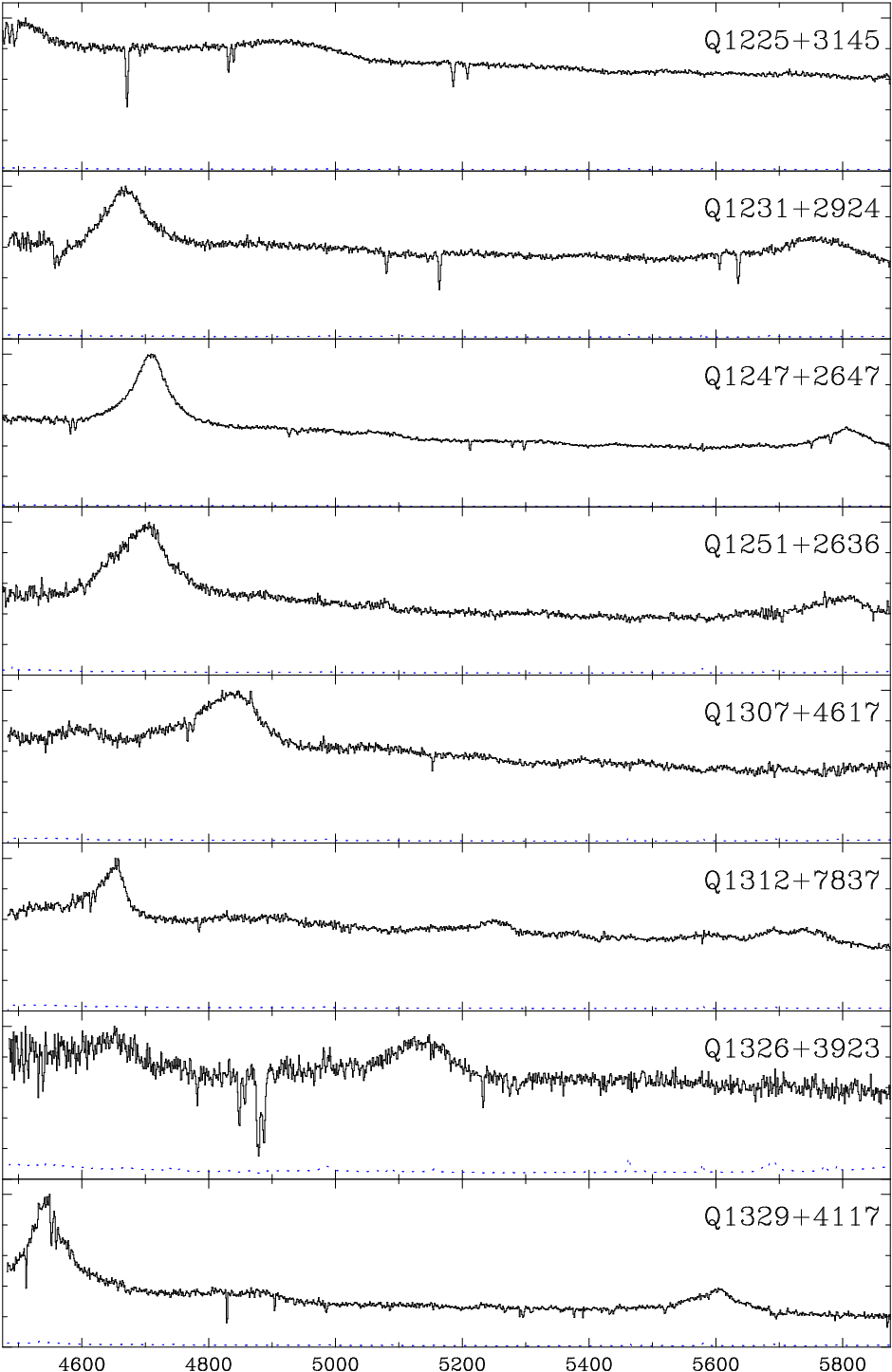}
\caption{Same as for Figure \ref{redspecfig}.}
\end{figure}

\clearpage

\begin{figure}
\epsscale{0.8}
\plotone{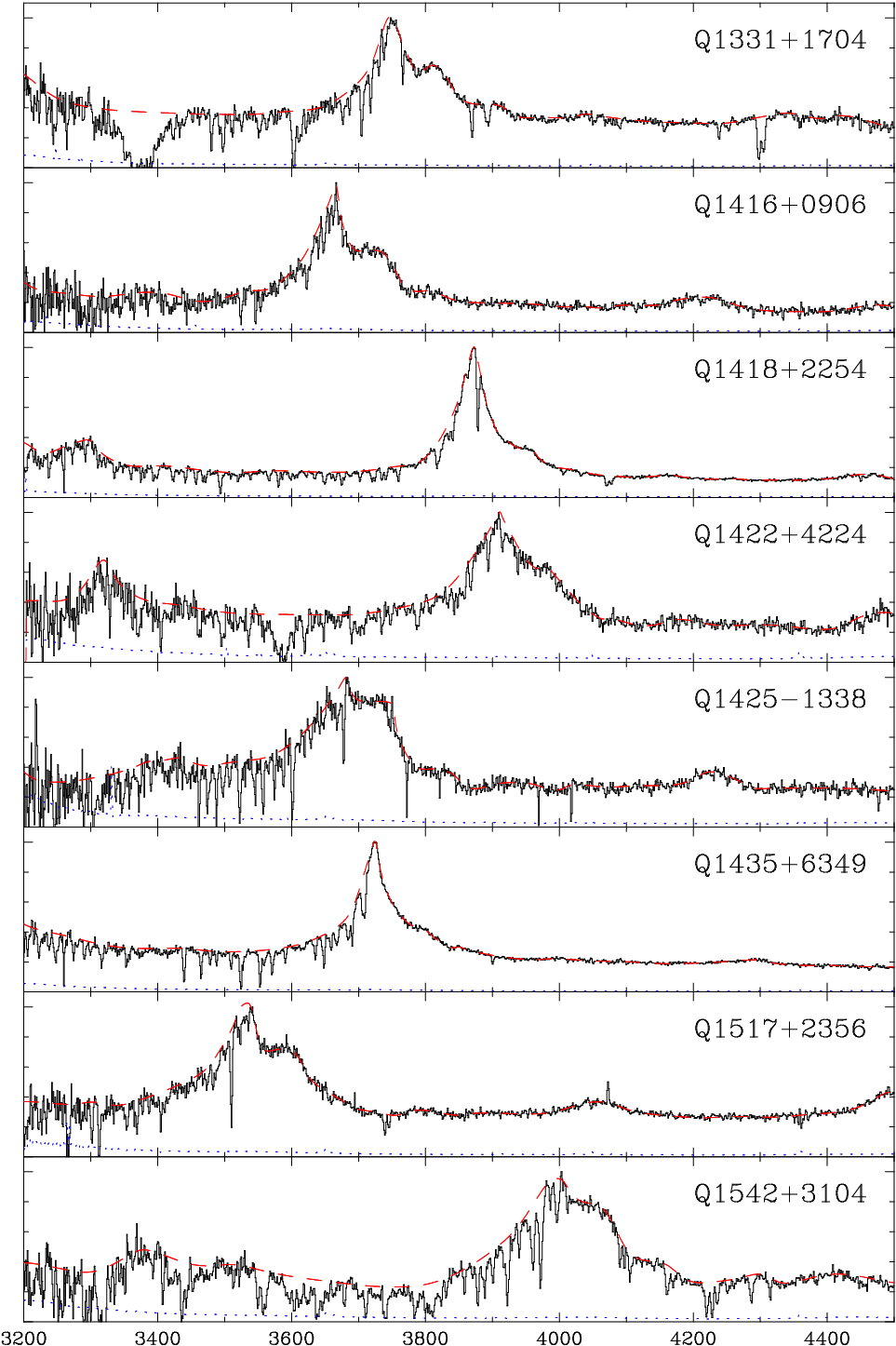}
\caption{Same as for Figure \ref{specfig}.}
\end{figure}

\begin{figure}
\epsscale{0.8}
\plotone{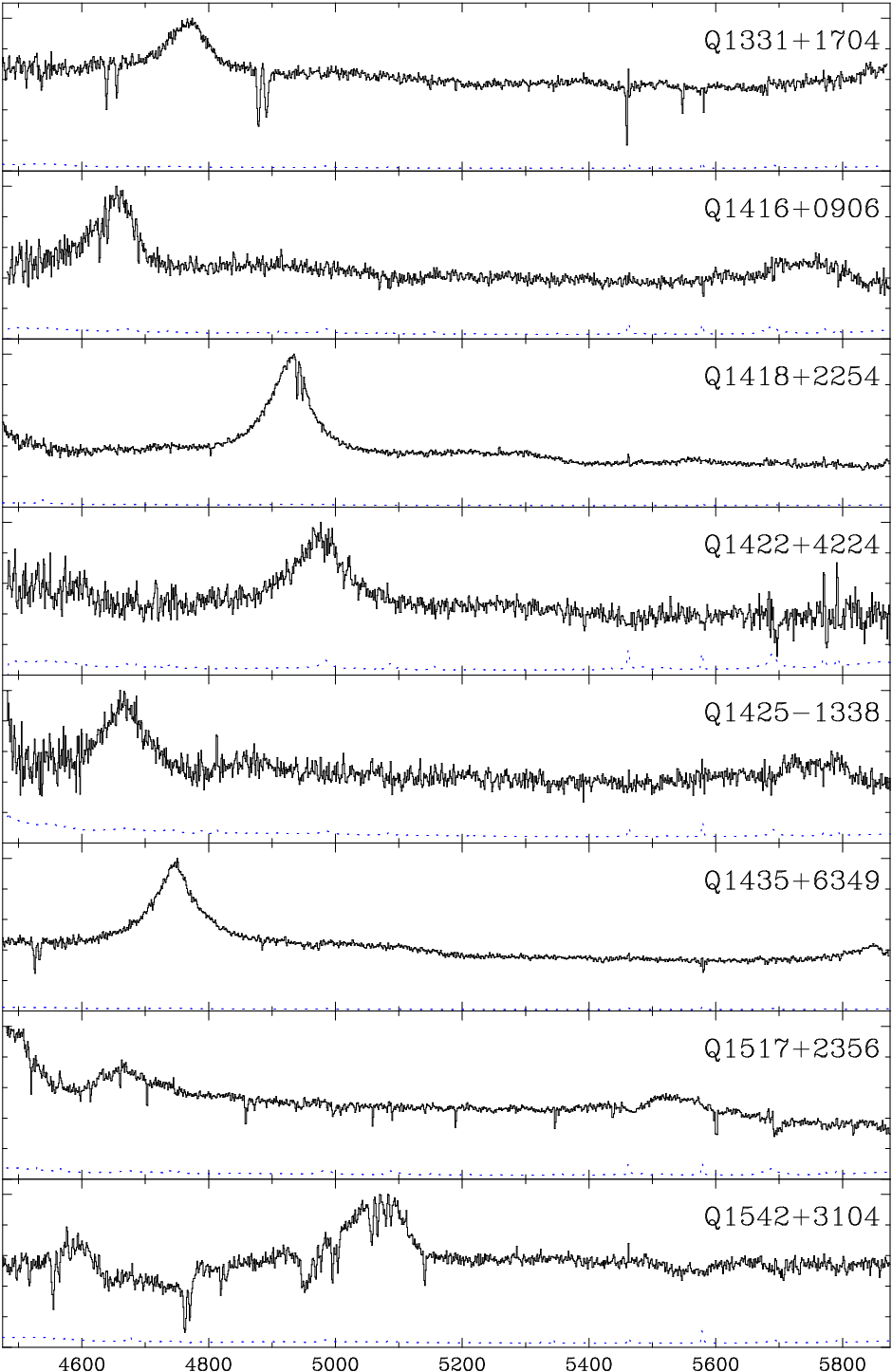}
\caption{Same as for Figure \ref{redspecfig}.}
\end{figure}

\clearpage

\begin{figure}
\epsscale{0.8}
\plotone{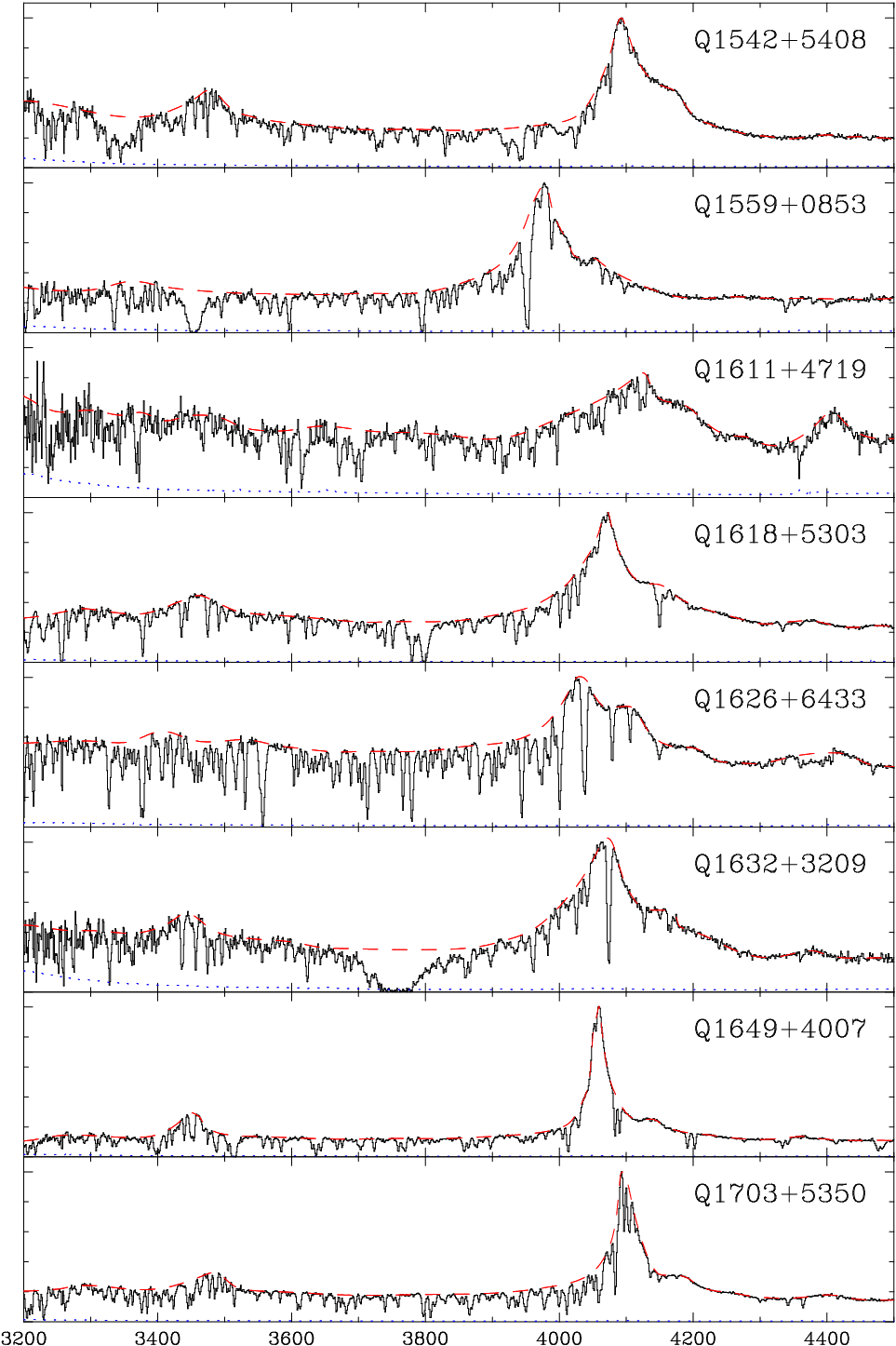}
\caption{Same as for Figure \ref{specfig}.}
\end{figure}

\begin{figure}
\epsscale{0.8}
\plotone{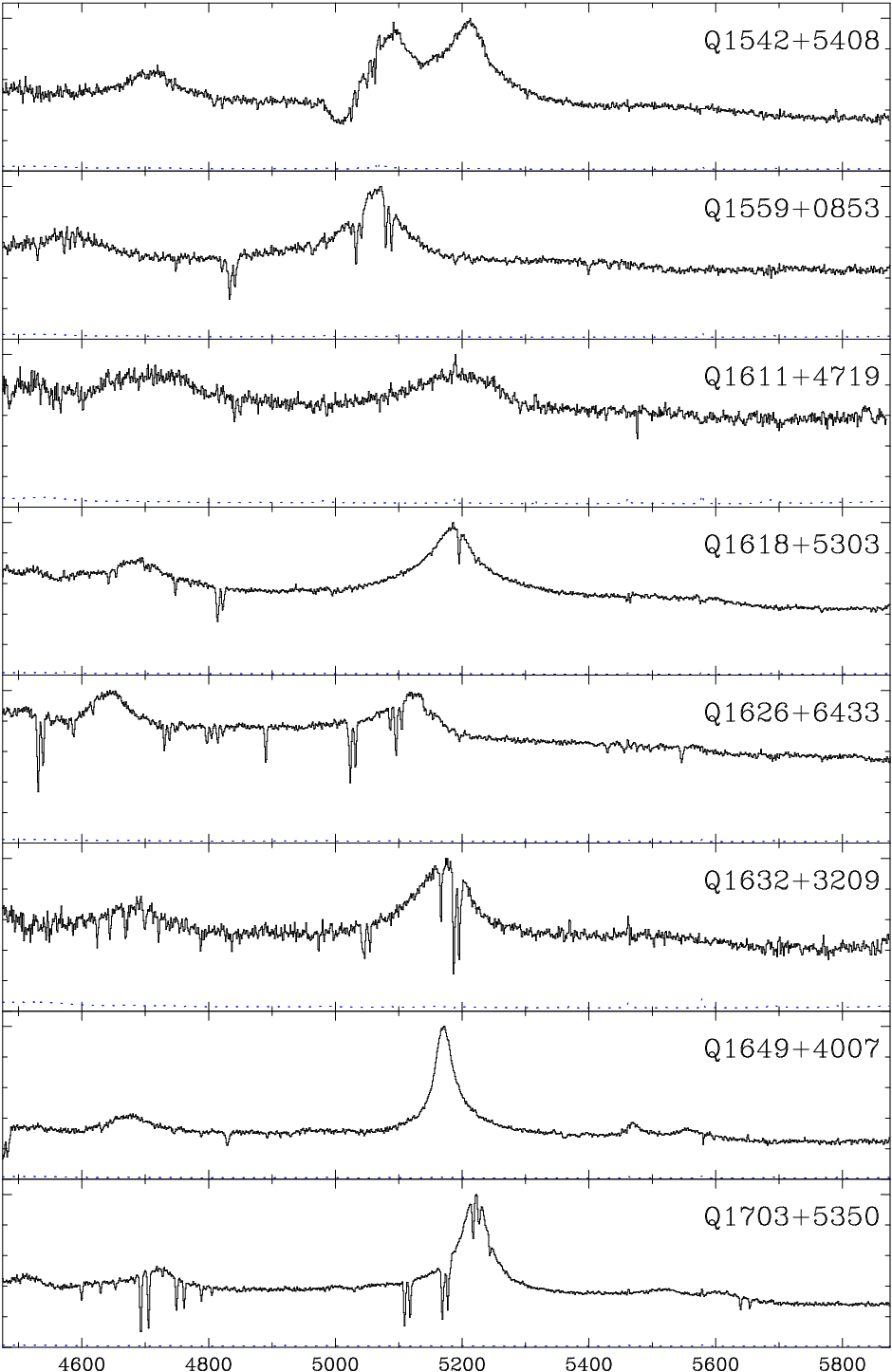}
\caption{Same as for Figure \ref{redspecfig}.}
\end{figure}

\clearpage

\begin{figure}
\epsscale{0.8}
\plotone{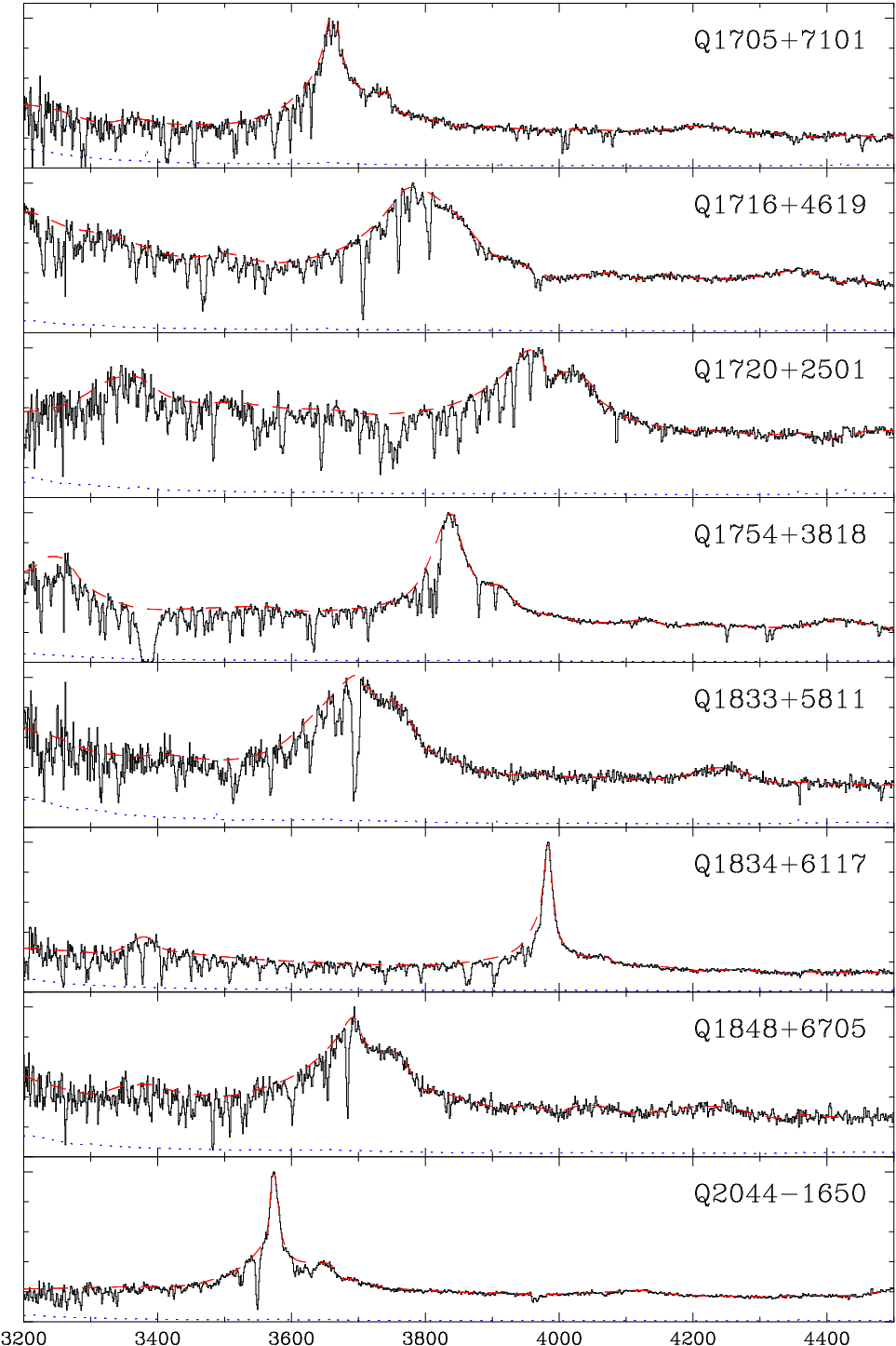}
\caption{Same as for Figure \ref{specfig}.}
\end{figure}

\begin{figure}
\epsscale{0.8}
\plotone{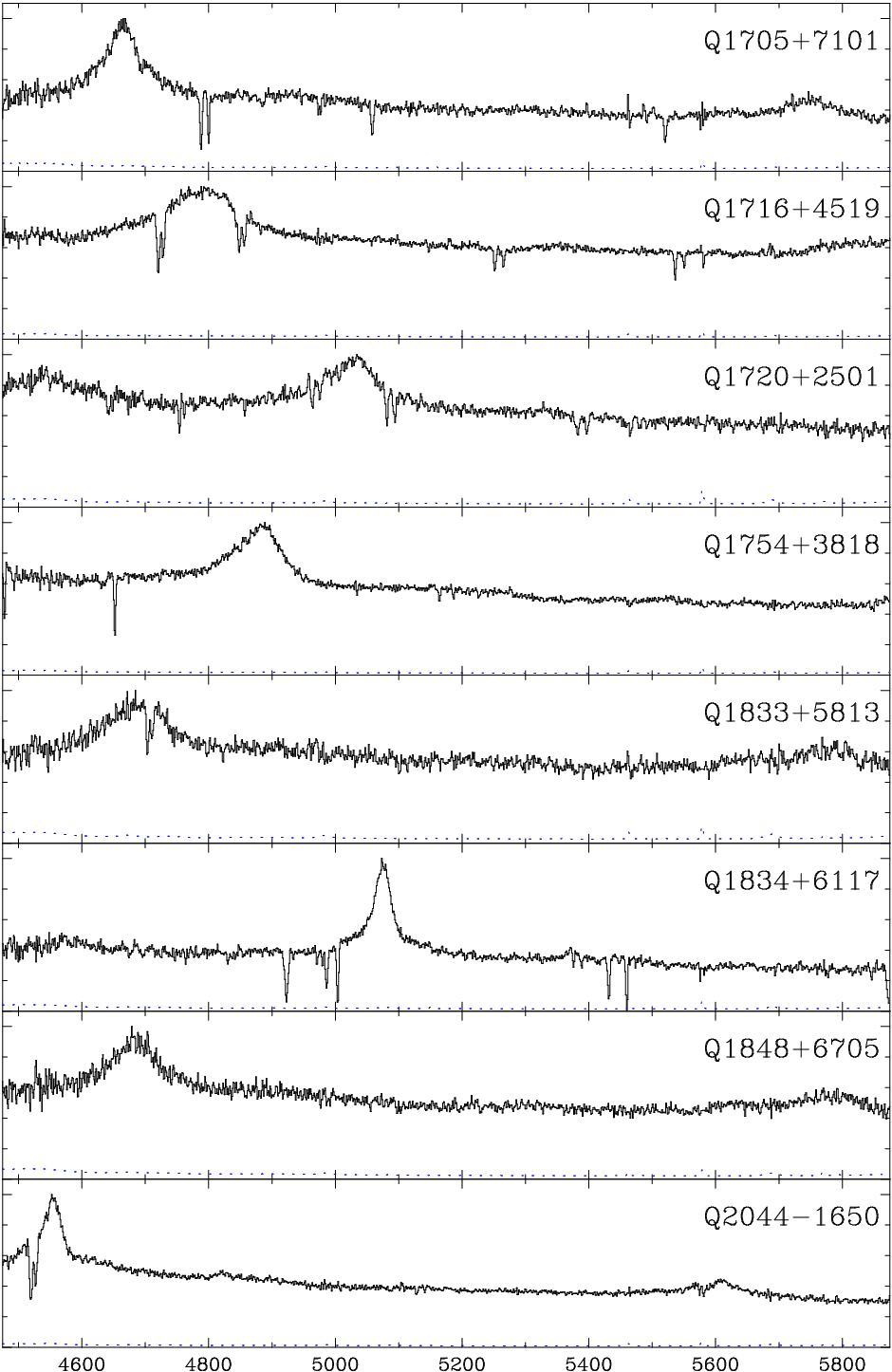}
\caption{Same as for Figure \ref{redspecfig}.}
\end{figure}

\clearpage

\begin{figure}
\epsscale{0.8}
\plotone{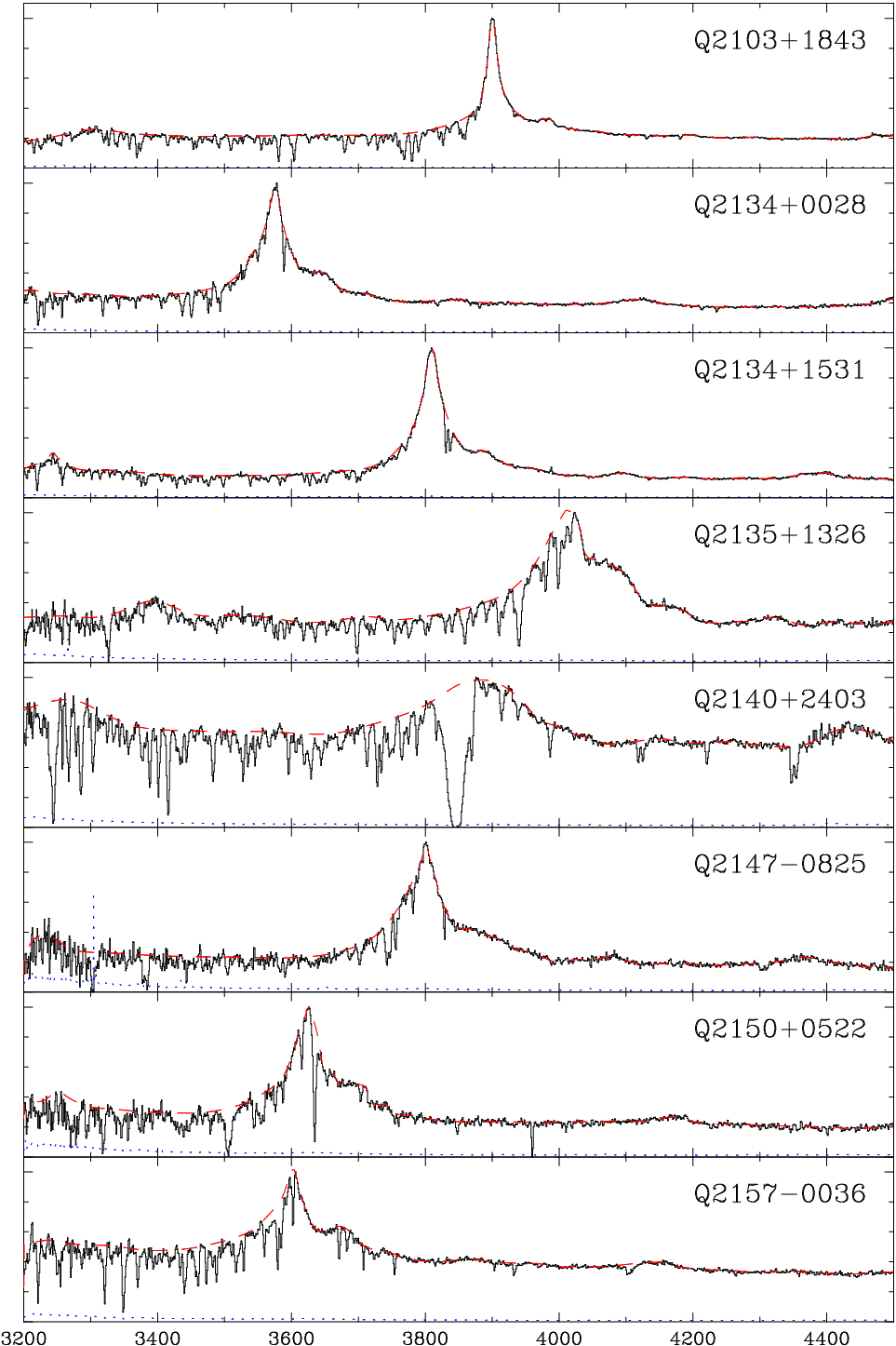}
\caption{Same as for Figure \ref{specfig}.}
\end{figure}

\begin{figure}
\epsscale{0.8}
\plotone{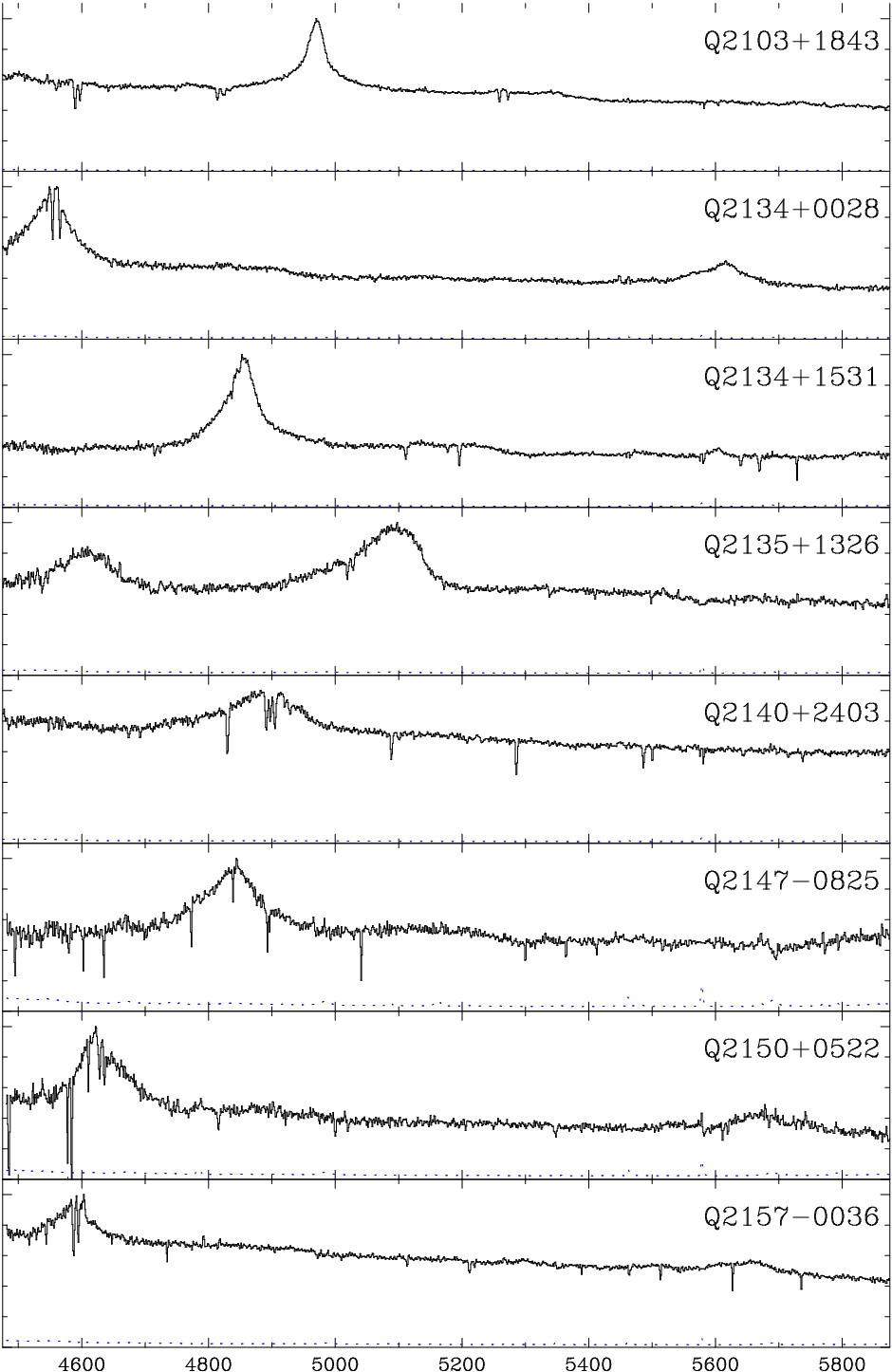}
\caption{Same as for Figure \ref{redspecfig}.}
\end{figure}

\clearpage

\end{document}